\documentclass[12pt]{article}
\usepackage{amsmath}
\usepackage{graphicx}
\usepackage{natbib}
\usepackage{url} % not crucial - just used below for the URL 
\usepackage{graphicx}
\usepackage{epsfig}
\usepackage{amsmath}
\usepackage{amssymb}
\usepackage{caption}
\usepackage{color}
\usepackage{enumerate}
\usepackage{multicol}
\usepackage{float}
\usepackage{subfig}
\usepackage{appendix}
\usepackage{setspace}
\usepackage{amsthm}

\newcommand{\blind}{0}

\newcommand{\be}{\begin{equation}}
\newcommand{\ee}{\end{equation}}
\newcommand{\ba}{\begin{eqnarray}}
\newcommand{\ea}{\end{eqnarray}}
\newcommand{\bpm}{\begin{pmatrix}}
	\newcommand{\epm}{\end{pmatrix}}

\newcommand{\beq}{\begin{equation}}
\newcommand{\eeq}{\end{equation}}
\newcommand{\bea}{\begin{eqnarray}}
\newcommand{\eea}{\end{eqnarray}}
\newcommand{\p}{^{\prime}}
\newcommand{\bs}{\boldsymbol}

\newcommand{\nt}{\notag} 

\newcommand{\bsbeta}{\bs{\beta}}

\newcommand{\bsX}{\bs{X}}

\newcommand{\kappasq}{\kappa^2}

\usepackage{algorithm}
\usepackage[noend]{algpseudocode}
\usepackage{algorithmicx}
\def\BState{\State\hskip-\ALG@thistlm}

\begin{document}

\def\spacingset#1{\renewcommand{\baselinestretch}%
{#1}\small\normalsize} \spacingset{1}

%%%%%%%%%%%%%%%%%%%%%%%%%%%%%%%%%%%%%%%%%%%%%%%%%%%%%%%%%%%%%%%%%%%%%%%%%%%%%%

\if0\blind
{
  \title{\bf Privacy for Spatial Point Process Data}
  \author{Adam Walder \hspace{.2cm}\\
    Department of Statistics, Pennsylvania State University, University Park\\
    \\
    Ephraim M. Hanks \\
    Department of Statistics, Pennsylvania State University, University Park\\
    \\
	Aleksandra Slavkovi\'c \\
	Department of Statistics, Pennsylvania State University, University Park\\}
  \maketitle
} \fi

\if1\blind
{
  \bigskip
  \bigskip
  \bigskip
  \begin{center}
    {\LARGE\bf Title}
\end{center}
  \medskip
} \fi

\bigskip
\begin{abstract}
	In this work we develop methods for privatizing spatial location data, such as spatial locations of individual disease cases. We propose two novel Bayesian methods for generating synthetic location data based on log-Gaussian Cox processes (LGCPs). We show that conditional predictive ordinate (CPO) estimates can easily be obtained for point process data. We construct a novel risk metric that utilizes CPO estimates to evaluate individual disclosure risks. We adapt the propensity mean square error (\textit{pMSE}) data utility metric for LGCPs. We demonstrate that our synthesis methods offer an improved risk vs. utility balance in comparison to radial synthesis with a case study of Dr. John Snow's cholera outbreak data.  
\end{abstract}

\noindent%
\vfill

\newpage
\spacingset{1.5} % DON'T change the spacing!

%%%%%%%%%%%%%%%%%%%%%%%%%%%%%%%%%%%%%%%%%%%%%%%%%%%%%%%%%%
%%%%%% Main Sections %%%%%%%%%%%%%%%%%%%%%%%%%%%%%%%%%%%%%
%%%%%%%%%%%%%%%%%%%%%%%%%%%%%%%%%%%%%%%%%%%%%%%%%%%%%%%%%%
	
\section{Introduction}

	Individual disease case data offer scientists valuable information about the dynamics of on-going and past disease outbreaks. Due to privacy risks, disease data are often not made publicly available. Synthetic data sets offer reduced disclosure risks while preserving some of the scientific utility of the confidential data set \citep{wang2012multiple,quick2015bayesian}. In this work, we consider spatial disease case data: that is we consider data consisting of the spatial locations of each individual with a particular disease. We develop methods for privatizing this fundamental form of disease data. We propose two novel Bayesian methods for generating synthetic individual location data based on log-Gaussian Cox processes (LGCPs). We demonstrate that leave-one-out LGCP densities can easily be approximated by conditional predictive ordinates (CPOs). We propose a novel disclosure risk metric based on a leave-one-out intrusion scenario, which is quickly evaluated by our derived CPO estimates. We propose a model-based utility metric tailored to LGCPs built off the propensity mean square error (\textit{pMSE}) explored by \cite{snoke2018pmse} to assess the scientific quality of a synthetic data set. We motivate our work with Dr. John Snow's renowned cholera outbreak dataset, which is, notably, one of the only openly available datasets consisting of individual disease case locations.   
	
\begin{figure}[H]
		\centering
		\includegraphics[scale=0.8]{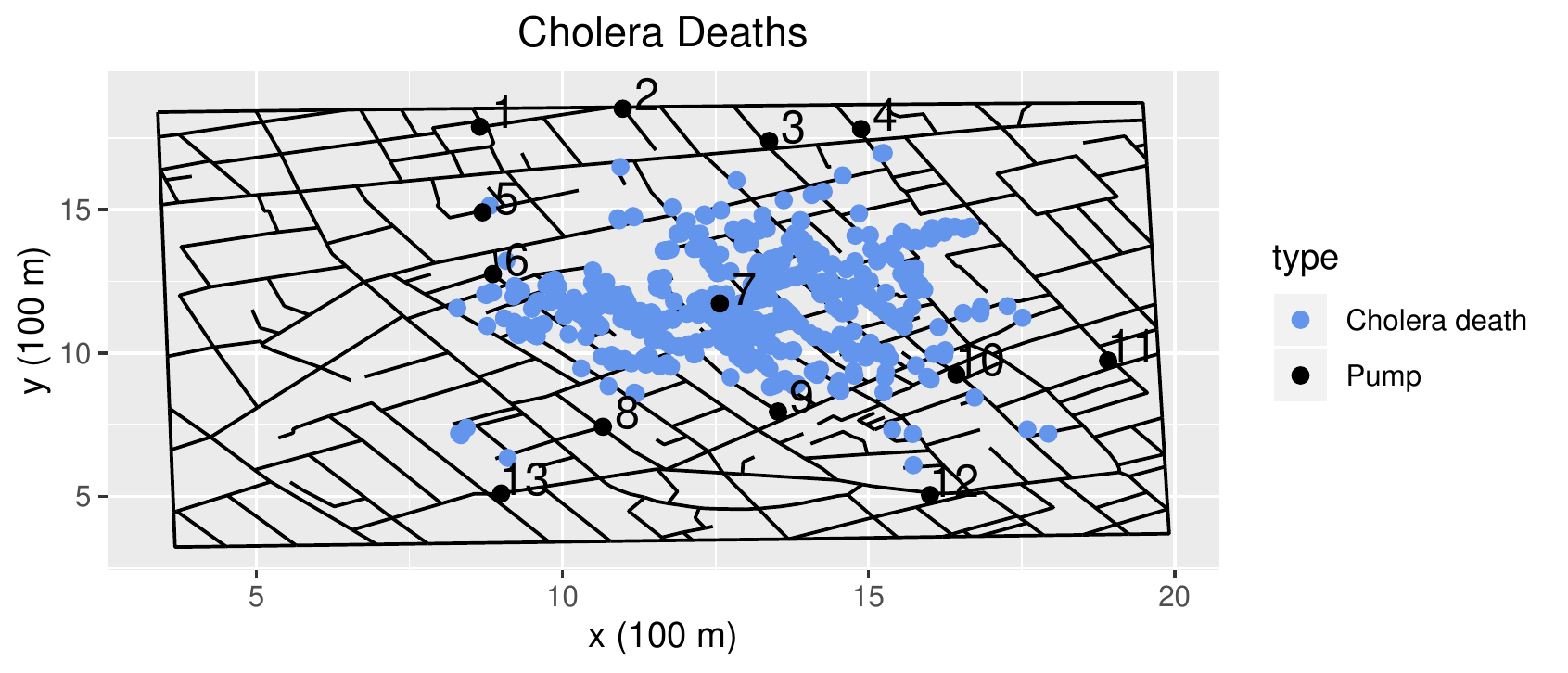}
		\caption{A map of the 578 observed cholera deaths, streets, and water pumps in Soho, London. The Broad St. water pump (pump 7) was the source of the outbreak.}
		\label{figure::Introduction Snow Cholera Deaths and Streets}
	\end{figure}

	In 1854, the small suburb of Soho, London was plagued by a massive cholera outbreak. When a wave of cholera first hit London in 1831, the scientific community at the time assumed that the disease was spread by ``miasma in the atmosphere" \citep{summers_1989}. Dr. John Snow, an obstetrician, sought to convince the town's officials that the epidemic was in fact water borne. To do so, Dr. Snow began by constructing a map of Soho (see Figure \ref{figure::Introduction Snow Cholera Deaths and Streets}). He then went door-to-door collecting, and plotting, each of the 578 reported cholera death incidences. From this map, Dr. Snow concluded that the Broad St. water pump was the source of the outbreak. This analysis was used to convince town officials to close the water pump.  
	
	Access to individual locations of each disease incidence allows for scientific analyses to accurately capture the geographic trends associated with a disease outbreak. Congress mandated the Health Insurance Portability and Accountability Act of 1996 (HIPAA) establishing standards for the privacy of individually identifiable health information \cite{aspe_2016}. For disease location data, individual identification is equivalent to knowing one's individual location. For this reason, agencies are legally obligated to ensure that proper privacy constraints are met before disseminating individual disease occurrence locations.

	One of the most common methods used to privatize disease case locations prior to data dissemination is radial perturbation \citep{armstrong1999geographically,wang2012multiple}. In radial perturbation, each location is randomly perturbed within a radius \textit{r} of their true location. It has been argued that radial perturbation poses two problems; 1) If perturbations of large radius \textit{r} are required for strong privacy guarantees, the underlying spatial structure is destroyed; 2) For sparsely populated regions, small perturbations offer weak privacy guarantees \citep{wang2012multiple,quick2015bayesian,quick2018generating}. Though the two stated drawbacks of radial perturbation seem to be agreed upon \citep{raghunathan2003multiple,wang2012multiple,quick2015bayesian}, these points have not been clearly illustrated by any empirical studies. In this work we demonstrate the fallacies of radial perturbation by providing the first Bayesian analysis of risk vs. utility of radially perturbed locations.

	Disease case locations which have been perturbed to satisfy legal constraints may be of low scientific quality. An ideal dataset will offer low disclosure risks and provide inference similar to that of the confidential dataset. \cite{wang2012multiple} proposed the use of CART models to generate synthetic locations from approximate conditional distributions of each location given the attributes of individuals in the data set. Since CART models draw locations independently from the response distribution they could produce a synthetic data set that contains two records that are spatially close in the confidential set but distant in the synthetic set \citep{wang2012multiple}. Although the CART models can preserve some spatial structure in the confidential data, they can miss localized spatial dependencies \citep{quick2015bayesian}. \cite{quick2015bayesian} used LGCPs in place of CART models to preserve the spatial dependence structure of the confidential set. \cite{quick2015bayesian} generated synthetic locations by simulating from the fitted LGCP model with posterior mean estimates for each parameter in the model. We propose a synthesis method that adds random spatial noise to the intensity surface of the LGCP. This method contains the synthesis method of \cite{quick2015bayesian} as a special case when the variance of the random spatial noise is set to zero. We also propose a second synthesis method that includes a resampled spatial random field in the intensity surface. We show that these novel synthesis methods provide improved disclosure risks relative to the methods of \cite{quick2015bayesian} and radial perturbation. 
	
	Synthetic data sets should offer quantifiable disclosure risks for every individual involved in the study. \cite{quick2015bayesian} defined a risk metric for fully-synthetic locations by conditioning on the attributes of each individual. This metric is not defined for location-only data sets, which are common for disease outbreaks and are the case we consider in this study. To assess disclosure risks we consider an intrusion scenario in which an intruder attempts to identify a confidential location within a radius \textit{r} of the truth given 1) knowledge of the synthesis method, 2) unique identification of all but one confidential location, and 3) the released synthetic dataset. In turn, our risk metric requires the evaluation of a leave-one-out posterior predictive density. Leave-one-out scenarios are commonly used to assess privacy risks and are central to privacy \citep{wang2012multiple, dwork2014algorithmic}. 
	
	Obtaining leave-one-out predictive densities is often a computationally burdensome task for high dimensional data sets and spatial models, as individual model fits are required for every observation in the data set. CPO estimates rely on independent marginal distributions to estimate the leave-one-out posterior predictive density for each observation by Monte Carlo approximation using samples from the posterior density of the full data set \citep{hooten2015guide}. In turn, CPO estimates are easy to obtain when independence is assumed for each observation. In the existing literature, dependence structures commonly assumed in spatial models make CPO estimates computationally expensive \citep{hooten2015guide}. In Section \ref{section::Conditional Predictive Ordinate Estimates for Cox Processes} we derive CPO estimates for Cox processes with dependence structures, allowing for fast approximations of posterior predictive leave-one-out LGCP densities for spatial point process data. This provides a general approach for model selection for spatial point process data, and in Section \ref{section::Disclosure Risk Metric} we demonstrate that each individual disclosure risk can be easily approximated with CPO estimates using samples from the posterior conditioned on the full data set. This provides a novel computationally efficient approach to assessing disclosure risks in spatial location data. 
	
	Synthetic data sets should satisfy disclosure risk requirements and offer meaningful scientific inference. \cite{woo2009global} and \cite{snoke2018general} developed the propensity mean square error (\textit{pMSE}) to assess the distributional similarity between a synthetic and confidential data set. The \textit{pMSE} is the mean square error of the predicted probability that a given observation belongs to the confidential set. A synthetic data set that is indistinguishable from the confidential set is said to be of high utility. We propose a model-based utility metric that tailors the \textit{pMSE} statistic to LGCPs. 
	
	In a case study of Dr. John Snow's cholera outbreak dataset, we demonstrate that both of our proposed synthesis methods offer datasets with reduced disclosure risks and higher data utility than radial perturbation. To our knowledge, we offer the first Bayesian analysis of radial perturbation. We also show that our synthesis methods offer improved disclosure risks relative to the approach of \cite{quick2015bayesian}. In summary, our contributions in this work include
	\begin{enumerate}
		\item Two novel methods for generating synthetic point process data based on log-Gaussian Cox processes (LGCPs).
		\item A derivation of conditional predictive ordinate (CPO) estimates for point process data.
		\item A novel metric for evaluating privacy risks based on CPO estimates.
		\item An adaptation of the propensity mean square error (\textit{pMSE}) data utility metric tailored to point process data. 
	\end{enumerate} 	
	
	The remainder of this manuscript is organized as follows. In Section \ref{section::Overview} we outline the process of generating and disseminating synthetic locations. In Section \ref{section::Log-Gaussian Cox Processes} we provide background information on the LGCPs models considered in this work. In Section \ref{section::Conditional Predictive Ordinate Estimates for Cox Processes} we demonstrate that the CPO can be readily obtained for any Cox process. In Section \ref{section::Generating Fully Synthetic Location Data}--\ref{section::Evaluating Utility} we introduce our proposed synthesis methodologies and detail our model based utility metric. In Section \ref{section::Evaluating Disclosure Risks} we introduce our novel risk metric and detail how individual disclosure risks are assessed for each synthesis method using CPO arguments. In Section \ref{section::Data Analysis} we illustrate our methodology with a case study of Dr. John Snow's cholera dataset. We conclude with a discussion in Section \ref{section::Discussion}.

\section{Data Synthesis and Dissemination} \label{section::Overview} 

	In this section we outline the process of generating, evaluating, and disseminating synthetic location-only data. We begin by fitting the confidential dataset according to a LGCP with intensity surface $\lambda(\cdot)$ as detailed in Section \ref{section::Log-Gaussian Cox Processes}. The LGCP allows for a model that includes any desired covariates and captures spatial auto-correlation.
	
	Once the model is fitted, synthetic locations are simulated from a LGCP with intensity $\lambda^{\dagger}(\cdot)$, determined from the confidential model fit. The LGCP intensity surface, $\lambda(\cdot)$, includes a zero-mean Gaussian spatial random effect with covariance parameterized by $\bs{\theta} = (\kappa^2,\tau^2)$. Our first proposed synthesis method re-samples the spatial random effect from a zero-mean Gaussian distribution with covariance determined by plug-in estimates $\hat{ \bs{\theta} }= (\hat{\kappa}^2,\hat{ \tau }^2)$. Our second novel synthesis method involves adding random Gaussian noise to the posterior mean estimate of the summary statistic $\hat{\lambda}(\cdot)$ from the confidential set. Synthetic locations are obtained by sampling from a LGCP with synthetic intensity $\lambda^{\dagger}(\cdot)$. 
	
	Prior to disseminating data, statistical agencies are required to ensure that subjects' can not be unwillingly identified \cite{aspe_2016}. We introduce an individual disclosure risk metric to quantify how much an intruder can learn about the confidential locations from a synthetic data set. We assess individual disclosure risks by considering the intrusion scenario in which the intruder has gained knowledge of 1) the synthetic dataset $\mathcal{S}^{\dagger}$, 2) complete information about the synthesis method, and 3) has uniquely identified all but one confidential location $\mathcal{S} / \bs{s}_k$. The intruder attempts to identify the final confidential location within a radius \textit{r} of the truth given the synthetic dataset and all but one confidential location. The risk is taken to be the max individual disclosure risk for $\mathcal{S}/\bs{s}_k$ given $\mathcal{S}^{\dagger}$ and complete information about the synthesis method. To evaluate this risk, we integrate the marginal density $\pi{\left( \bs{s},N | \mathcal{S}/\bs{s}_k, \mathcal{S}^{\dagger}\right)}$ over a ball of radius \textit{r} centered at confidential location $\bs{s}_k$. We require samples from the posterior distributions of $\pi{\left(\lambda |\mathcal{S}^{\dagger},\mathcal{S} / \bs{s}_k, N \right)}$ to evaluate the risks for each of the \textit{N} confidential locations in the Bayesian framework. We use samples from the posterior distribution obtained from a simultaneous fit of all the data, $\pi{\left( \lambda, \lambda^{\dagger} | \mathcal{S}^{\dagger}, \mathcal{S},N\right)}$, to estimate the CPO $\pi{\left(\bs{s},N |\mathcal{S}^{\dagger},\mathcal{S} / \bs{s}_k\right)}$. We demonstrate that CPO estimates can be used to approximate the disclosure risks for each synthesis method considered in Section \ref{section::Evaluating Disclosure Risks}.  	
	
	The goal of this work is to provide approaches to disseminate fully synthetic location-only data that offer low disclosure risks and provide scientific inference similar to that of the confidential set. Once a synthetic dataset is generated, we assess the quality (utility) of a synthetic dataset by computing the \textit{pMSE} statistic to quantify how well a given synthetic dataset emulates the confidential dataset. We then evaluate the maximum disclosure risk for our given intrusion scenario. An optimal synthetic dataset will provide a small maximum disclosure risk, while maintaining high data utility. We repeat the processes of data generation and risk/utility assessment until an acceptable balance of data utility vs. maximum disclosure risk has been obtained. A synthetic dataset that satisfies the desired privacy vs. utility trade-off is then released along with all information related to the synthesis process.

\section{LGCPs} \label{section::Log-Gaussian Cox Processes} 
	In this section we provide a brief background on LGCPs and the computational details associated with model fitting. A Cox process is a point process governed by a non-negative stochastic process $\Lambda = \{\bs{s}\in\mathbb{R}^2: \Lambda(\bs{s}) \}$. Conditioned on the realization $\Lambda(\bs{s}) = \lambda(\bs{s})$, the point process is a Poisson process with intensity $\lambda(\bs{s})$. Cox processes are natural models for point process phenomena that are environmentally driven, such as the spatial locations of infectious disease cases \citep{diggle2013spatial}. In this work, we model locations according to a Cox process with a spatially varying intensity surface $\lambda(\cdot)$. The number of points inside a region $\Omega \subset \mathbb{R}^2$ is distributed $N|\lambda(\cdot) \sim Pois\left(\int_{\Omega} \lambda(\bs{s})d\bs{s}\right)$. The likelihood for a set of locations $\mathcal{S} = \{\bs{s}_i\}_i^N$ observed in $\Omega$ is given by 
	\bea 
	\pi(\mathcal{S},N|\lambda(\cdot)) = \exp\left(-\int_{\Omega} \lambda(\bs{s})d\bs{s} \right) \frac{\prod_{i=1}^{N}\lambda(\bs{s}_i)}{N!}. \label{Eqn::LGCP Cox Process Likelihood}
	\eea 
	
	\cite{moller1998log} introduced the class of LGCPs as a method to describe spatial correlation in point process models. A LGCP with spatially continuous covariates $\bs{x}(\bs{s})$ and population density offset $\log{(pd(\bs{s}))}$ is a Cox process with intensity $\lambda(\bs{s})$ given by 
	\bea 
		\log(\lambda(\bs{s})) = \log( pd(\bs{s}) ) + \bs{x}\p(\bs{s})\bsbeta + \eta(\bs{s}), \quad \bs{s}\in \mathbb{R}^2,  \label{Eqn::LGCP Log Lambda Intensity}
	\eea 
	where $\eta(\bs{s})$ is a Gaussian process with mean zero and some user defined covariance function $C(\cdot,\cdot)$. 
	
	\subsection{Computational Details for Fitting Log-Gaussian Cox Processes} \label{section::Computational Details for Fitting Log-Gaussian Cox Processes}
	Fitting LGCPs is a computationally burdensome task for high dimensional data. In this work we elect to use the approximation technique of \cite{lindgren2011explicit} to express the Gaussian process $\eta(\bs{s})$ as a basis expansion 
	\bea 
		\eta(\bs{s}) = \sum_{i=1}^{n} \phi_i(\bs{s}) w_i, \quad \bs{s} \in \Omega, \label{Eqn::LGCP Basis Expansion}
	\eea  
	where $n$ is the number of knots placed in $\Omega$ and $\{\phi_i(\bs{s})\}_{i=1}^{n}$ is a set of piecewise triangular basis functions (see Appendix \ref{Appendix::Finite Element Approximations for Matern GRFs} for details). \cite{lindgren2011explicit} showed that the weights of the basis expansion in \eqref{Eqn::LGCP Basis Expansion} are distributed 
	\bea 
		\bs{w}| \kappasq,\xi^2 \sim N(\bs{0}, \xi^2 Q^{-1}_{\kappasq}), \label{Eqn::LGCP Basis Weight Distribution} 
	\eea 
	where $Q^{-1}_{\kappasq}$ is a sparse precision matrix. The distribution in \eqref{Eqn::LGCP Basis Weight Distribution} is an approximation to a zero-mean Gaussian process with Mat\'ern covariance function 
	\bea 
		C(\bs{u},\bs{v}) = \xi^2 (\kappa||\bs{v}-\bs{u}||)K_1(\kappa||\bs{v}-\bs{u}||), \quad \bs{v},\bs{u} \in \mathbb{R}^2, \label{Eqn::LGCP Matern Covariance}
	\eea  
	where $||\cdot||$ denotes Euclidean distance, and $K_1(\cdot)$ is an order one Bessel function of the second kind. The marginal variance is given by $\sigma^2 = \xi^2 / \left(4\pi \kappa^2\right)$, while the approximation of the effective range is given by $\rho = \sqrt{8}/\kappa$ \citep{lindgren2011explicit}. We refer the reader to Appendix (\ref{Appendix::Finite Element Approximations for Matern GRFs}--\ref{Appendix::Approximate Inference for LGCPs}) for further details regarding model fitting.

	\section{Conditional Predictive Ordinate Estimates for Cox Processes} \label{section::Conditional Predictive Ordinate Estimates for Cox Processes}  	

	\cite{geisser1980discussion} first introduced the leave-one-out predictive distribution $\pi{\left( y_i|\bs{y}_{-i} \right)}$, known as the conditional predictive ordinate (CPO), as a diagnostic to detect inconsistent observations from a given model. CPO estimates are commonly used to perform Bayesian model selection for independently distributed error responses, as only one model fit is required to obtain each leave-one-out-predictive density \citep{pettit1990conditional,hooten2015guide}. CPO estimates are obtained by Monte Carlo approximation, using samples from the posterior distribution conditioned on the full data set \citep{hooten2015guide}. In this section we demonstrate that CPO estimates can easily be obtained for Cox processes with dependence structures. 
	
	The leave-one-out predictive distribution for each location in a given dataset is
	\bea 
		CPO_k = \pi{\left( \bs{s}_k, N | \mathcal{S}_{-k} \right) } = \int \pi{ \left( \bs{s}_k,N|\lambda,\mathcal{S}_{-k} \right) }\pi{\left( \lambda | \mathcal{S}_{-k} \right)} d\lambda. \label{Eqn::CPO CPO Defn} 
	\eea 	
	The quantity in \eqref{Eqn::CPO CPO Defn}, also known as the conditional predictive ordinate (CPO), represents the density of $\bs{s}_k$ when a model is fit without $\bs{s}_k$ \citep{geisser1980discussion}. CPO estimates are constructed by utilizing independent marginal distributions to rewrite the likelihood without observation $k$, $\pi{\left(\mathcal{S}_{-k}|\lambda,N \right) }$, as a scaled factor of the full likelihood $\pi{\left(\mathcal{S}|\lambda,N \right) }$. When $\bs{s}_k$ is independent of $\mathcal{S}_{-k}$ (i.e. independence), samples from the full posterior $\pi{\left( \lambda | N, \mathcal{S} \right)}$ can be used to easily obtain CPO estimates via Monte Carlo approximation of \eqref{Eqn::CPO CPO Defn}. The statistic $-2\sum_{i=1}^{N} CPO_i$ is commonly used to perform Bayesian model selection in a fashion similar to leave-one-out cross-validation \citep{hooten2015guide}.  
	
	Dependence structures commonly assumed in spatial models are generally computationally expensive, and so CPOs are rarely used for spatial data. Here, we demonstrate that CPO estimates can easily be obtained for Cox processes with spatial dependence structures. The key reason this is possible is that, for a LGCP, the likelihood of each location is independent of all other locations conditioned on the intensity surface $\lambda(\cdot)$, which contains the dependence structure. In Section \ref{section::Evaluating Disclosure Risks}, we define a risk metric based on a leave-one-out intrusion scenario. The CPO estimates derived in this section allow us to quickly obtain disclosure risk estimates by avoiding the computationally burdensome task of leave-one-out model fitting.
	
	Assume that the points, $\mathcal{S} = \{\bs{s}_i\}_{i=1}^{N}$, follow a Cox Process with intensity $\lambda(\cdot)$. $CPO_k$ is given by
	\bea 
		CPO_k = \pi(\bs{s}_k,N | \mathcal{S}_{-k}) = \left(\mathbb{E}_{ \pi{ \left( \lambda | \mathcal{S},N \right) } }\left[ \frac{\int_{\Omega}\lambda(\bs{s})}{\lambda(\bs{s}_k)} \right] \right)^{-1}.	\label{Eqn::CPO}	
	\eea   
	To see this, we first note that we can express the likelihood $\pi(\mathcal{S}_{-k},N|\lambda)$ as 
	\bea
		\pi(\mathcal{S}_{-k},N|\lambda) &=& \int_{\Omega} \frac{\prod_{i=1}^{N} \lambda(\bs{s}_i)}{N! \exp{\left( \int_{\Omega} \lambda(\bs{s}) d\bs{s} \right)}} d\bs{s}_k \nt \\ 
		&=&  \frac{\left( \int_{\Omega} \lambda(\bs{s})d\bs{s} \right) \left( \prod_{i \neq k} \lambda(\bs{s}_i) \right) }{N! \exp{\left( \int_{\Omega} \lambda(\bs{s}) d\bs{s} \right)}} \nt \\
		&=& \frac{\lambda(\bs{s}_k)}{\int_{\Omega} \lambda(\bs{s})d\bs{s}} \pi(\mathcal{S},N|\bs{\theta}). \label{Eqn::CPO Proportionality}
	\eea 
	Using \eqref{Eqn::CPO Proportionality}, we obtain the CPO as follows
	\bea 
		CPO_k  = \pi(\bs{s}_k,N| \mathcal{S}_{-k}) &=& \left(\frac{ \pi(\mathcal{S}_{-k},N) }{\pi(\mathcal{S},N)}\right)^{-1} \nt \\
		&=& \left(\int \frac{ \pi(\mathcal{S}_{-k},N| \lambda ) }{\pi(\mathcal{S},N)} \pi(\lambda) d\lambda \right)^{-1} \nt \\
		&=& \left(\int \left( \frac{ \int_{\Omega} \lambda(\bs{s})d\bs{s} }{\lambda(\bs{s}_k)} \right) \frac{ \pi(\mathcal{S},N| \lambda ) }{\pi(\mathcal{S},N)} \pi(\lambda) d\lambda \right)^{-1} \nt \\
		&=& \left(\int \left( \frac{ \int_{\Omega} \lambda(\bs{s})d\bs{s} }{\lambda(\bs{s}_k)} \right) \pi(\lambda | \mathcal{S}, N ) d\lambda \right)^{-1} \nt \\
		&=& \left(\mathbb{E}_{\pi{ \left( \lambda | \mathcal{S}, N\right)} }\left[ \left( \frac{ \int_{\Omega} \lambda(\bs{s})d\bs{s} }{\lambda(\bs{s}_k)} \right) \right] \right)^{-1}. \nt 
	\eea 
	
	We have shown that the CPO can easily be expressed in terms of expectation with respect to the posterior conditioned on the full data set $\pi{\left( \lambda|\mathcal{S},N \right) }$ for any Cox process. Thus CPO estimates are easily be obtained by a Monte Carlo approximation for spatial point process data generated by LGCPs! To see this, let $\lambda^{(m)}$ represent the $m^{th}$ sample drawn from $\pi{\left( \lambda|\mathcal{S},N \right) }$. CPO estimates are obtained by 
	\bea 
		CPO_k = \pi(\bs{s}_k,N| \mathcal{S}_{-k}) \approx \left( \sum_{m=1}^{M}\frac{ \int_{\Omega} \lambda^{(m)}(\bs{s})d\bs{s} }{\lambda^{(m)}(\bs{s}_k)} \right)^{-1}, \label{Eqn::CPO CPO Estimate} 
	\eea 
	where $M$ in \eqref{Eqn::CPO CPO Estimate} represents the total number of samples drawn from $\pi{\left( \lambda|\mathcal{S},N \right) }$. In Section \ref{section::Evaluating Disclosure Risks}, we demonstrate that individual disclosure risks can be easily evaluated by CPO estimates.

\section{Generating Fully Synthetic Location Data} \label{section::Generating Fully Synthetic Location Data}
	In this section we detail our proposed Bayesian methods for generating fully synthetic location data based on LGCPs. We also formally introduce radial synthesis as a baseline comparison for our proposed synthesis methods.

	\subsection{Radial Synthesis} \label{section::Radial Synthesis}
	Randomly perturbing locations within a radius \textit{r} of their true location is one of the most common redaction methods \citep{vanwey2005confidentiality}. It has been claimed that when large perturbations are required to ensure adequate privacy protection, inference related to spatial dependence structures are diminished \citep{armstrong1999geographically,wang2012multiple,quick2015bayesian}. Intuitively this claim seems reasonable, as large perturbations will destroy localized spatial clusters. Due to its popularity and simplicity, we treat radial perturbation as the baseline synthesis method for comparison. To our knowledge, no Bayesian analyses have been performed to assess the utility and disclosure risks associated with radial perturbation. We perform radial synthesis by drawing $N$ independent synthetic locations from circular uniform distributions with radius $r$ centered at each $\bs{s}_i$, denoted $\bs{s}_i^{\dagger} \sim U(\bs{s}_i,r)$.
	
	\subsection{Additive Noise Synthesis} \label{section::Additive Noise Synthesis} 
	Here, we introduce our first proposed synthesis method, Additive Noise Synthesis (\textit{ANS}). In \textit{ANS}, we alter the global variance parameter of the spatial random field $\eta(\bs{s})$ contained within the intensity surface $\lambda(\bs{s})$. We perform \textit{ANS} by adding a noisy spatial random field $\nu(\bs{s})$ with the same spatial scale to the log-intensity surface. \textit{ANS} proceeds as follows. We first fit an LGCP to the confidential set $\mathcal{S} = \{\bs{s}_i\}_i^N$ according to Section \ref{section::Log-Gaussian Cox Processes}. We then obtain posterior mean estimates for the basis expansion weights $\hat{\bs{w}} = \mathbb{E}_{ \pi{ \left( \bs{w},\bs{\beta}|\mathcal{S} \right) } }[\bs{w}]$, fixed effects $\hat{\bsbeta} = \mathbb{E}_{ \pi{ \left( \bs{w},\bs{\beta}|\mathcal{S} \right) } }[\bsbeta]$, and spatial scale parameter $\hat{\kappa}^2 = \mathbb{E}_{ \pi{\left( \kappasq,\xi^2|\mathcal{S} \right)} }[\kappasq]$. 
	
	Next, we simulate a Gaussian noise process $\bs{v} \sim N(\bs{0}, \sigma^2 \bs{Q}^{-1}_{\widehat{\kappa}^2})$ with precision matrix $\bs{Q}^{-1}_{\kappasq}$ (defined in Appendix \ref{Appendix::Finite Element Approximations for Matern GRFs}) for some user-defined noise level $\sigma^2$. Following Section \ref{section::Computational Details for Fitting Log-Gaussian Cox Processes}, we express the additive noise as a basis expansion $\nu(\bs{s}) = \sum_{i=1}^n \phi_i(\bs{s})v_i$. We obtain the resulting \textit{ANS} intensity surface 
	\bea 
		\lambda^{\dagger}(\bs{s}) = \exp\left( \log(pd(\bs{s}))+\bs{x}\p(\bs{s}) \hat{\bsbeta} + \sum_{i=1}^{n} \phi_i(\bs{s}) \left( \hat{w}_i+v_i \right) \right). \label{Eqn::ANS Intensity Surface}
	\eea 
	Note that $\bs{w}$ and $\bs{v}$ are both mean zero normally distributed random vectors with covariance structures given by $\xi^2 \bs{Q}^{-1}_{\kappasq}$ and $\sigma^2 \bs{Q}^{-1}_{{\kappa}^2}$ and thus, $\bs{w}+\bs{v} \sim N(\bs{0}, (\xi^2+\sigma^2) \bs{Q}^{-1}_{{\kappa}^2})$. Intuitively, the addition of Gaussian random noise in the intensity surface will randomly scale the relative risk of disease occurrence at each location $\bs{s}$. To see this note that \eqref{Eqn::ANS Intensity Surface} can equivalently be expressed as $\lambda^{\dagger}(\bs{s})  = e^{ \nu(\bs{s})} \hat{\lambda}(\bs{s})$, where $\hat{\lambda}(\bs{s})$ denotes the intensity surface with plug-in posterior mean estimates $\hat{\bsbeta}$ and $\hat{\bs{w}}$. We also note that \textit{ANS} can be viewed as an extension of the synthesis method proposed by Quick et al. \cite{quick2015bayesian}, in which synthetic locations were generated based on posterior mean plug-in estimates. We obtain the synthesis method of \cite{quick2015bayesian} by simply taking the noise-level to be zero ($\sigma^2=0$).  
	
	We generate synthetic locations by uniformly sampling $N^*$ points with $N^* >> N$ over the spatial domain $\Omega$. We then assign each of the $N^{*}$ sampled locations $\{\bs{s}^*_i\}_{i=1}^{N^*}$ probability weights $\hat{p}_k^* = \frac{\lambda^{\dagger}(\bs{s}_k^*)}{\sum_{j=1}^{N^*} \lambda^{\dagger}(\bs{s}_j^*)}$. We obtain an \textit{ANS} synthetic dataset by sampling \textit{N} of the $N^*$ locations without replacement according to their probabilities $\hat{p}^{*}_k$.	
	
	\begin{figure}[H]
		\centering
		\includegraphics[scale=0.75]{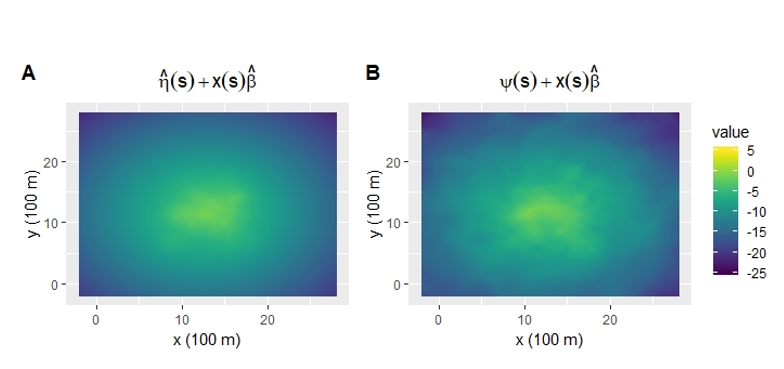}
		\caption{(A): An intensity surface plot with the posterior mean estimate for the spatial random field $\hat{\eta}(\bs{s})$ from the John Snow cholera outbreak dataset. (B): An \textit{ANS} intensity surface plot for an additive noise spatial random field, $\psi(\bs{s})$, with spatial scale $\hat{\kappa}^2$ and noise-level $\sigma^2=10$. }
		\label{figure::Additive_Noise_Synthesis ANS Eta}
	\end{figure}
	
	In \textit{ANS}, we alter the marginal variance of the spatial random effect by some user defined noise level $\sigma^2$. Therefore, an \textit{ANS} data set produces an intensity surface with the same spatial scale as the confidential data set with larger variability (see Figure \ref{figure::Additive_Noise_Synthesis ANS Eta}). In turn, we expect \textit{ANS} data sets to offer an inverse relationship between disclosure risks and data utility relative to the user defined noise level. That is, an \textit{ANS} data set with a large user defined noise level will offer lower disclosure risks and lower data utility relative to a data set with a smaller noise level.    
	
	\subsection{Posterior Resampling Synthesis} \label{section::Posterior Resampling Synthesis}
\begin{figure}[H]
		\centering
		\includegraphics[scale=0.75]{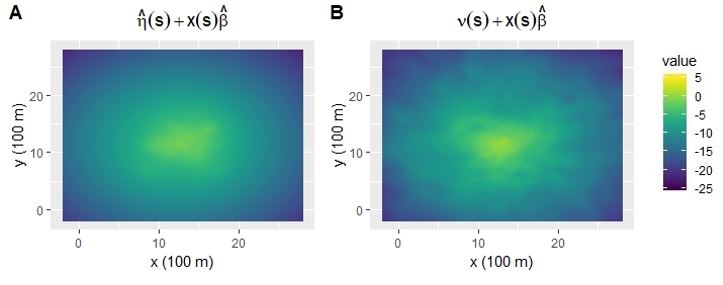}
		\caption{(A): An intensity surface plot with the posterior mean estimate for the spatial random field $\hat{\eta}(\bs{s})$ from the John Snow cholera outbreak dataset. (B): An intensity surface plot for a resampled spatial random field, $\nu(\bs{s})$, with spatial scale $\hat{\kappa}^2$ and variance parameter $\hat{\xi}^2$. }
		\label{figure::Posterior Resampled Synthesis Resampled Eta}
	\end{figure}

	Here, we propose another novel synthesis method which we call Posterior Resampling Synthesis (\textit{PRS}). Gaussian processes with Mat\'ern covariance given by \eqref{Eqn::LGCP Matern Covariance} are fully characterized by variance parameter $\xi^2$ and spatial scale parameter $\kappasq$. For fixed values of $\xi^2$ and $\kappasq$, realizations of the spatial random field $\eta(\bs{s})$ will produce random fields with the same marginal variance and spatial range. Intuitively, a resampled spatial random field will relocate spatial clusters while preserving the dependence structure of the confidential intensity surface (see Figure \ref{figure::Posterior Resampled Synthesis Resampled Eta}).  
	
	The posterior resampling synthesis (\textit{PRS}) process is described as follows. Similar to \textit{ANS}, we first fit the confidential dataset according to a LGCP as described in Section \ref{section::Computational Details for Fitting Log-Gaussian Cox Processes} via MCMC. We then compute posterior mean estimates for the fixed effects, $\hat{\bsbeta} = \mathbb{E}_{ \pi{ \left(\bsbeta,\bs{w}|\mathcal{S} \right) }}[\bsbeta]$, spatial scale, $\hat{\kappa}^2 = \mathbb{E}_{\pi{ \left( \xi^2,\kappasq| \mathcal{S} \right) }}[\kappasq]$, and variance parameter, $\hat{\xi}^2 = \mathbb{E}_{\pi{ \left( \xi^2,\kappasq| \mathcal{S} \right) }}[\xi^2]$ from the confidential fit. We generate a new spatial random field based on the posterior mean of the hyperparameters, $\bs{w}^{*} \sim N(\bs{0}, \hat{\xi}^2 Q^{-1}_{\hat{\kappa}^2})$. We obtain the \textit{PRS} intensity surface by swapping the basis expansion weights $\bs{w}$ for the resampled weights $\bs{w}^{*}$, 
	\bea 
		\lambda^{\dagger}(\bs{s}) = \exp{ \left( \log(pd(\bs{s})) + \bs{x}\p(\bs{s})\bsbeta + \sum_{i=1}^{n} \phi_i(\bs{s} )w_i^{*} \right) }. \label{Eqn::PRS Intensity}
	\eea 
	We then simulate a \textit{PRS} dataset by sampling $N$ locations from a LGCP with intensity \eqref{Eqn::PRS Intensity} as described in Section \ref{section::Additive Noise Synthesis}. 
	
	The spatial scale and marginal variance of a \textit{PRS} intensity surface are given by posterior mean estimates from the confidential data set. In turn, \textit{PRS} generated data sets maintain data utility by preserving the correlative structure of the confidential set, and reduce disclosure risks by relocating spatial clusters contained in the intensity surface. 

\section{Evaluating Utility} \label{section::Evaluating Utility} 

	In Sections (\ref{section::Radial Synthesis}--\ref{section::Posterior Resampling Synthesis}) we presented radial synthesis and our two proposed methods for generating fully synthetic location data. Our goal is to disseminate synthetic locations that reduce disclosure risks, while allowing for meaningful analyses to be conducted by secondary analysts. The propensity mean square error (\textit{pMSE}) developed by \cite{woo2009global} and \cite{snoke2018general} is the mean square error of the predicted probability that a given observation belongs to the confidential set. The \textit{pMSE} is a metric for measuring the distributional similarity between a synthetic and confidential data set \citep{snoke2018pmse}. In this section, we provide a model-based estimation of the predicted probabilities required to estimate the \textit{pMSE} for LGCPs.
	
	Let $\lambda(\bs{s})$ and $\lambda^{\dagger}(\bs{s})$ denote the intensity surfaces for the true dataset, $\mathcal{S}$, and synthetic dataset, $\mathcal{S}^{\dagger}$, respectively. Let $\{\bs{y}_k \}_{k=1}^{2N}$ denote the collection of true and synthetic locations $\mathcal{S} \cup \mathcal{S}^{\dagger}$. Assuming $P(\bs{y}_k \in \mathcal{S}) = 0.5$, the probability a point $\bs{y}_k$ belongs to the true dataset is
	\bea 
		p_k = P(\bs{y}_k \in \mathcal{S} | \lambda, \lambda^{\dagger} ) = \frac{\lambda(\bs{y}_k) }{\lambda(\bs{y}_k) + \left(\frac{\int_{\Omega} \lambda(\bs{s})d\bs{s} }{ \int_{\Omega} \lambda^{(\dagger)}(\bs{s})d\bs{s}  } \right)\lambda^{\dagger}(\bs{y}_k)}.  \label{Eqn::Utility Pk}
	\eea

	The probability of correct classification for each point in $\mathcal{S} \cup \mathcal{S}^{\dagger}$ can be approximated by Monte Carlo simulation. To do so, \textit{L} samples are drawn from the intensity surface of the true posterior $\lambda^{(l)} \sim \pi{ \left( \lambda | \mathcal{S} \right) }$ and the synthetic posterior ${\lambda^{\dagger}}^{(l)} \sim \pi{ \left( \lambda^{\dagger} | \mathcal{S}^{\dagger} \right) }$ respectively. Define
	\bea 
		p_k^{(l)} = P(\bs{y}_k \in \mathcal{S} | \lambda^{(l)}, {\lambda^{\dagger}}^{(l)} ) = \frac{\lambda^{(l)}(\bs{y}_k) }{\lambda^{(l)}(\bs{y}_k) + \left(\frac{\int_{\Omega} \lambda^{(l)}(\bs{s})d\bs{s} }{ \int_{\Omega} {\lambda^{\dagger}}^{(l)}(\bs{s})d\bs{s}  } \right) {\lambda^{\dagger}}^{(l)} (\bs{y}_k)}. \label{Eqn::Utility Pk_l}
	\eea 
	The probability of correctly identifying $\bs{y}_k$ is 
	\bea 
		\hat{p}_k = E[P(\bs{y}_k \text{ is classified correctly})] \approx \frac{1}{L} \sum_{l=1}^{L} (p_k^{(l)})^{ x_k } (1-p_k^{(l)})^{1-x_k}, \label{Eqn::Utility Pk hat}
	\eea 
	where
	\[x_k = \begin{cases} 
	0, & \bs{y}_k \in \mathcal{S}^{\dagger} \\
	1, & \bs{y}_k \in \mathcal{S} 
	\end{cases}. \]
	
	The predicted probabilities given in \eqref{Eqn::Utility Pk hat} are used to to approximate the \textit{pMSE}, 
	\bea 
		pMSE \approx \widehat{pMSE} =  \frac{1}{2N} \sum_{k=1}^{2N} (\hat{p}_k - 0.5)^2  \label{Utility_pMSE}.
	\eea 	
	We elect to use the \textit{pMSE} to quantify the quality of a given synthetic data set due to its intuitive interpretation. A synthetic dataset, $\mathcal{S}^{\dagger}$, that is indistinguishable from the confidential set, $\mathcal{S}$, produces a \textit{pMSE} score of 0. A synthetic dataset that is systematically distinguishable from the confidential set will offer little scientific inference. Such sets are assigned a worst case \textit{pMSE} score of 0.25.
	
	In this section, we have demonstrated how the \textit{pMSE} can be tailored to LGCPs. In Section \ref{section::Data Analysis}, we use the \textit{pMSE} to assess the quality of synthetic cholera death locations.

\section{Evaluating Disclosure Risks} \label{section::Evaluating Disclosure Risks}
	In this section we construct a novel risk metric to quantify individual disclosure risks in spatial location data. We provide the computational details required to obtain individual disclosure risk estimates for each proposed synthesis method. We also discuss the restrictive a priori assumptions required to obtain differentially private intensity surfaces $\lambda(\cdot)$.
	
	\subsection{Disclosure Risk Metric} \label{section::Disclosure Risk Metric} 
	Let $\mathcal{S} = \{\bs{s}_i\}_{i=1}^{N}$ be the collection of the \textit{N} confidential locations and $\mathcal{S}^{\dagger} = \{\bs{s}^{\dagger}_i\}_{i=1}^{N}$ the collection of \textit{N} synthetic locations. Assume an adversary has access to the synthetic dataset, knowledge of the synthesis method, population density, and has identified all but the $k^{th}$ confidential location $\bs{s}_k$. The intruder aims to identify the final $k^{th}$ confidential location within a ball of radius \textit{r} of the truth, denoted $\mathcal{B}_{r}(\bs{s}_k)$. We consider this high risk scenario, as it quantifies how much information a synthetic set offers to an intruder who has identified all but one true location. This is a common scenario in privacy, including empirical differential privacy \citep{wang2012multiple,charest2016meaning}. The metric also holds an intuitive interpretation; if a region $\mathcal{B}_r(\bs{s}_k)$ is of high probability then the synthetic data set provides an intruder with information useful for locating a confidential location.   
	
	The probability of identifying person $k$ within a radius \textit{r} of the true location given the synthetic dataset and all but the $k^{th}$ true location is 
	\bea 
		P( \{  \bs{s} \in \mathcal{B}_{r}(\bs{s}_k) \cap \Omega \} | \mathcal{S}_{-k},\mathcal{S}^{\dagger} ) =  \int_{ \mathcal{B}_{r}(\bs{s}_k) } \pi{\left(\bs{s}|\mathcal{S}_{-k},\mathcal{S}^{\dagger} \right)} d\bs{s}. \label{Eqn::Risk Integral Marginal sk}
	\eea 
	Notice that \eqref{Eqn::Risk Integral Marginal sk} is an integral over a leave-one-out-density $\pi{ \left( \bs{s} | \mathcal{S}_{-k},\mathcal{S}^{\dagger}  \right) }$. We rely on the CPO estimates detailed in Section \ref{section::Conditional Predictive Ordinate Estimates for Cox Processes} along with a quadrature scheme (detailed in Appendix \ref{Appendix::Quadrature Scheme For Circular Domains}) to evaluate \eqref{Eqn::Risk Integral Marginal sk}. The computational details required for evaluating the CPOs, $\pi{\left(\bs{s}|\mathcal{S}_{-k},\mathcal{S}^{\dagger} \right)}$, of each synthesis method are described in the following sections. 
	
	\subsection{Radial Synthesis Disclosure Risks} \label{section::Radial Synthesis Disclosure Risks} 
	To fit a LGCP to a radially perturbed synthetic data set and the confidential set simultaneously, we assume a circular uniform prior for the synthetic data, $\bs{s}_i^{\dagger} \sim U(\bs{s}_i,r)$ for $i=1,2,...,N$, and a LGCP with intensity $\lambda(\cdot)$ for the confidential set $\mathcal{S}$. We require an estimate of the CPO $\pi{ \left( \bs{s} | \mathcal{S}_{-k},\mathcal{S}^{\dagger}  \right) }$ to evaluate the risk metric in \eqref{Eqn::Risk Integral Marginal sk}. Evaluation of the leave-one-out density $\pi{ \left( \bs{s} | \mathcal{S}_{-k},\mathcal{S}^{\dagger}  \right) }$ requires samples from the posterior distribution $\pi{ \left( \lambda | \mathcal{S}_{-k},\mathcal{S}^{\dagger}  \right) }$ for each location $\bs{s}_k$. We also require a linkage prior to determine which synthetic location $\bs{s}_i^{\dagger}$ was uniquely generated from each confidential location $\bs{s}_i$. 
	
	Fortunately we can overcome this computational bottleneck by augmenting the CPO argument presented in Section \ref{section::Conditional Predictive Ordinate Estimates for Cox Processes}. The leave-one-out predictive density for data generated by radial synthesis is given by  
	\bea 
		\pi{ \left( \bs{s},N | \mathcal{S}_{-k}, \mathcal{S}^{\dagger} \right) }  = \left[ \mathbb{E}_{ \pi{\left(\lambda | \bs{S} \right)} }\left[ \frac{\int_{\mathcal{B}_r(\bs{s}_k^{\dagger})} \lambda(\bs{s})d\bs{s}}{\lambda(\bs{s})} \right] \right]^{-1}. \label{Eqn::Radial Synthesis CPO}
	\eea 
	Notice that the CPO estimate in \eqref{Eqn::Radial Synthesis CPO} only requires samples from the marginal distribution of the confidential dataset $\pi{ \left( \lambda|\mathcal{S} \right) }$. In turn, we do not require samples drawn from the leave-one-out posterior density $\pi{\left( \lambda | \mathcal{S}_{-k},\mathcal{S}^{\dagger} \right)}$. We also avoid the computationally burdensome task of forming a linkage prior between the confidential and synthetic locations. A full derivation of \eqref{Eqn::Radial Synthesis CPO} is included in Appendix \ref{Appendix::Circular Synthesis Disclosure Risk Details}.   
	
	We estimate the CPO in \eqref{Eqn::Radial Synthesis CPO} by Monte Carlo integration. Let $L$ represent the number of samples drawn from the marginal density $\pi{ \left( \lambda | \mathcal{S} \right) }$. The expectation in \eqref{Eqn::Radial Synthesis CPO} is estimated by 
	\bea 
		\pi{ \left( \bs{s}, N | \mathcal{S}_{-k}, \mathcal{S}^{\dagger} \right) } \approx \left[ \frac{1}{L} \sum_{l=1}^{L} \frac{ \int_{\mathcal{B}_r(\bs{s}_k^{\dagger})} \lambda^{(l)}(\bs{s})d\bs{s} }{ \lambda^{(l)}(\bs{s}) }\right]^{-1}. \label{Eqn::Radial MC Approx}	
	\eea 
	Note that the integral $\int_{\mathcal{B}_r(\bs{s}_k^{\dagger})} \lambda^{(l)}(\bs{s})d\bs{s}$ in \eqref{Eqn::Radial MC Approx} must be evaluated for all $L$ samples. We approximate the analytically intractable integral with the quadrature scheme detailed in Appendix \ref{Appendix::Quadrature Scheme For Circular Domains}. 
	
	\subsection{Additive Noise Synthesis Disclosure Risks} \label{section::Additive Noise Synthesis Disclosure Risks}
	CPO estimates of $\pi{\left( \bs{s} | \mathcal{S}_{-k}, \mathcal{S}^{\dagger} \right) }$ for \textit{ANS} generated sets require samples from the joint distribution $\pi{  \left( \lambda,\lambda^{\dagger} | \mathcal{S}, \mathcal{S}^{\dagger} \right) }$. We use a Metropolis Hastings sampler to draw samples from the desired joint marginal density. The simultaneous fit assumes that the synthetic data is independent of the confidential data, that is $ \pi{ \left( \mathcal{S},\mathcal{S}^{\dagger} | \lambda, \lambda^{\dagger} \right) } = \pi{ \left( \mathcal{S} | \lambda \right) } \pi{ \left( \mathcal{S}^{\dagger} | \lambda^{\dagger} \right) }$, where $\pi{ \left( \mathcal{S} | \lambda \right) } \sim LGCP(\lambda(\cdot))$ and $\pi{ \left( \mathcal{S}^{\dagger} | \lambda^{\dagger} \right)} \sim LGCP\left( \lambda^{\dagger}(\cdot)  \right)$. The intensity surfaces are of the form 
	\bea 
	\log{ \left( \lambda(\bs{s}) \right) } &=& \bs{x}'(\bs{s})\bsbeta + \sum_{i=1}^n \phi_i(\bs{s}) w_i + \log{ \left( pd\left( \bs{s} \right) \right)  } \label{Eqn::ANS Risk True Lam} \\
	\log{ \left( \lambda^{\dagger}(\bs{s}) \right) } &=& \bs{x}'(\bs{s})\bsbeta + \sum_{i=1}^n \phi_i(\bs{s}) (w_i+v_i) + \log{ \left( pd\left( \bs{s} \right) \right)  }, \label{Eqn::ANS Risk Syn Lam} 
	\eea 
	where the basis expansion weights in \eqref{Eqn::ANS Risk True Lam} and \eqref{Eqn::ANS Risk Syn Lam} are independently distributed $\pi{ \left( \bs{w} | \xi^2,\kappasq \right) } \sim N(\bs{0}, \xi^2 Q^{-1}_{\kappasq})$ and $\pi{ \left( \bs{v} | \sigma^2, \kappasq \right) } \sim N(\bs{0},\sigma^2 Q^{-1}_{\kappasq})$.
	Recall that $\sigma^2$ is the noise level, released to the public along with the synthetic dataset. For this reason $\sigma^2$ is treated as known and fixed at its true value. 
	
	Let $\lambda^{(l)}(\cdot)$ denote the $l^{th}$ draw from $\pi{ \left(\lambda,\lambda^{\dagger} | \mathcal{S},\mathcal{S}^{\dagger} \right) }$. A Monte Carlo estimate of the CPO, $\pi{ \left( \bs{s} | \mathcal{S}_{-k}, \mathcal{S}^{\dagger} \right) }$, is given by  
	\bea 
	\pi{ \left( \bs{s}, N | \mathcal{S}_{-k}, \mathcal{S}^{\dagger} \right) } = \left[ \mathbb{E}_{ \pi{ \left(\lambda,\lambda^{\dagger}|\mathcal{S},\mathcal{S}^{\dagger} \right) } }\left[ \frac{ \int_{\Omega} \lambda(\bs{s})d\bs{s} } {\lambda(\bs{s})} \right] \right]^{-1}  \approx \left[ \frac{1}{L} \sum_{l=1}^L \frac{ \int_{\Omega} \lambda^{(l)}(\bs{s})d\bs{s} } {\lambda^{(l)}(\bs{s})} \right]^{-1}. \label{Eqn::ANS Risk MC Approx}		
	\eea   
	The integral $\int_{\Omega} \lambda^{(l)}(\bs{s})d\bs{s}$ in \eqref{Eqn::ANS Risk MC Approx} is numerically approximated for all $L$ samples following \cite{simpson2016going} (see \eqref{Eqn::Appendix Voronoi Dual Mesh Sum} of Appendix \ref{Appendix::Approximate Inference for LGCPs} for details). The individual disclosure risk in \eqref{Eqn::Risk Integral Marginal sk} for location $\bs{s}_k$ is obtained by using a quadrature scheme (see Appendix \ref{Appendix::Quadrature Scheme For Circular Domains}) to integrate the CPO over $\mathcal{B}_r{\left( \bs{s}_k \right) }$. 
	
	\subsection{Posterior Resampling Synthesis Disclosure Risks} \label{section::Resampled Synthesis Disclosure Risks} 
	Similar to \textit{ANS}, CPO estimates for \textit{PRS} require samples from the joint marginal $\pi{ \left( \lambda, \lambda^{\dagger} | \mathcal{S}, \mathcal{S}^{\dagger} \right) }$. As done with \textit{ANS}, we again assume that the synthetic data is independent of the confidential set, that is $\pi{ \left( \lambda, \lambda^{\dagger} | \mathcal{S},\mathcal{S}^{\dagger} \right) }= \pi{ \left( \lambda | \mathcal{S} \right) } \pi{ \left( \lambda^{\dagger} | \mathcal{S}^{\dagger} \right) }$. Recall that the synthetic intensity surface is given by $\log{\left( \lambda^{\dagger}(\bs{s}) \right) } = \bs{x}\p(\bs{s})\bsbeta + \sum_{i=1}^{n} \phi_i(\bs{s})w_i^{*} + \log{ ( pd(\bs{s}) ) }$, where $\bs{w}^{*} \sim N(\bs{0}, \hat{\xi}^2 Q^{-1}_{\hat{\kappa}^2} )$. 
	
	A Metropolis Hastings sampler is used to obtain samples from the joint distribution by sequentially drawing from the posterior distributions $\pi{ \left( \bsbeta | \mathcal{S},\mathcal{S}^{\dagger},\bs{w},\bs{w}^{*} \right) }$, $\pi{ \left( \bs{w} | \mathcal{S},\bsbeta,\kappasq,\xi^2 \right) }$, $\pi{ \left( \bs{w}^{*} | \mathcal{S}^{\dagger},\bsbeta,\kappasq,\xi^2 \right) }$ and $\pi{ \left( \xi^2, \kappasq | \bs{w},\bs{w}^{*} \right) }$. CPO estimates are obtained by evaluating \eqref{Eqn::ANS Risk MC Approx} using samples from the joint distribution. \textit{PRS} disclosure risks are then obtained by following the quadrature and Monte Carlo scheme as detailed in Section \ref{section::Additive Noise Synthesis Disclosure Risks}.

	\subsection{Differential Privacy} \label{section::Differntial Privacy} 
	
	\cite{dwork2006calibrating} and \cite{dwork2006differential} introduced differential privacy (DP) a measure of confidentiality protection. DP protects the information of every individual in the data set by limiting the influence that any one respondent can have on the released information. DP ensures the confidentiality of each individual in a database, even against an adversary who has gained complete knowledge of the rest of the data set. 
	
	Formally, a randomized algorithm $\mathcal{M}(\mathcal{S})$ is said to be $\epsilon$-differentially private if 
	\bea 
		P(\mathcal{M}(\mathcal{S}) \in A ) \leq \exp{(\epsilon)}P(\mathcal{M}(\mathcal{S}^{*}) \in A ), \label{Eqn::DP eps-DP}
	\eea 
	for all measurable subsets $A$ of the range of $\mathcal{M}$ and for all datasets $\mathcal{S},\mathcal{S}^{*}$ differing by one entry \citep{dwork2014algorithmic}. \cite{dimitrakakis2014robust} and \cite{foulds2016theory} observed that Bayesian posterior sampling provides $\epsilon$-differential privacy under certain prior assumptions. Theorem 2 of \cite{foulds2016theory} states that releasing one sample from the posterior distribution, $\pi{\left(\bs{x}|\theta \right)}$, with any prior, $\pi{\left( \theta \right)}$, is $2C$-differentially private provided $sup_{\bs{x},\bs{x}\p \in \mathcal{X}, \theta \in \Theta} | \log{\left( \pi{\left(\bs{x}|\theta \right)} \right)} | \leq C$. We describe this result for LGCPs. 
	
	Consider the collection of $N$ locations in $\Omega \subset \mathbb{R}^2$ denoted $\mathcal{X}$. Let $\mathcal{S},\mathcal{S}^{*} \in \mathcal{X}$ differ by at most one entry, say $\bs{s}$ and $\bs{s}^{*}$. Assume we model the locations according to a $LGCP$ with intensity $\lambda(\cdot)$ given by \eqref{Eqn::LGCP Log Lambda Intensity} with a spatial random field defined as in \eqref{Eqn::LGCP Basis Weight Distribution}. As shown in \cite{foulds2016theory}, releasing one sample from the posterior distribution $\pi{ \left( \lambda | \mathcal{S} \right) }$ is $2C$-differentially private for any prior, provided 
	\bea 
		max_{\substack{\mathcal{S},\mathcal{S}' \in \mathcal{X},\lambda \in \Lambda }} |\log{\pi{ \left( \bs{s}^{*} | \lambda \right)} } - \log{ \pi{ \left( \bs{s} | \lambda \right)} } | \leq C. \label{Eqn::DP Bayesian DP Def}
	\eea 
	For LGCPs, \eqref{Eqn::DP Bayesian DP Def} holds provided $\forall \bs{s},\bs{s}^{*} \in \Omega$, 
	\bea 
		sup_{\substack{\lambda \in \Lambda }} | \log{ \lambda{\left(\bs{s}^{*}\right)} } - \log{ \lambda(\bs{s}) } | \leq C.   \label{Eqn::DP Lam max diff}
	\eea 
	Equivalently, the left hand side of \eqref{Eqn::DP Lam max diff} can be expressed as
	\bea 
		sup_{\substack{\lambda \in \Lambda }} \left\vert \left(\bs{x}'\left( \bs{s}^{*} \right) - \bs{x}'(\bs{s})\right)\bsbeta  + \sum_{i=1}^{n} \left(\phi_i{\left(\bs{s}^{*}\right)} - \phi_i{(\bs{s})} \right)w_i + \log{\left( pd\left(\bs{s}^{*}\right) \right)} - \log{ \left(pd(\bs{s}) \right)} \right\vert. \label{Eqn::DP Bound} 
	\eea 
	It is clear from \eqref{Eqn::DP Bound} that the privacy ``cost" \textit{C} in \eqref{Eqn::DP Lam max diff} is determined by bounding the maximum distance between any continuous covariate measured on $\Omega$ and the magnitude of each fixed effect $\beta_i$ and spatial weight $w_i$. An a priori constraint on the parameter space of both the fixed effects and spatial random effects is required to produce a desired privacy cost \textit{C}. Scientists rarely posses enough a priori knowledge to suggest a simultaneous clipping of the fixed effects and spatial random effect. For this reason, we have elected to move away from differential privacy, and instead utilize the disclosure risk metric introduced in Section \ref{section::Disclosure Risk Metric}.

\section{Case Study: Dr. John Snow's Cholera Outbreak} \label{section::Data Analysis}
\begin{figure}[H]
		\centering
		\includegraphics[scale=0.5]{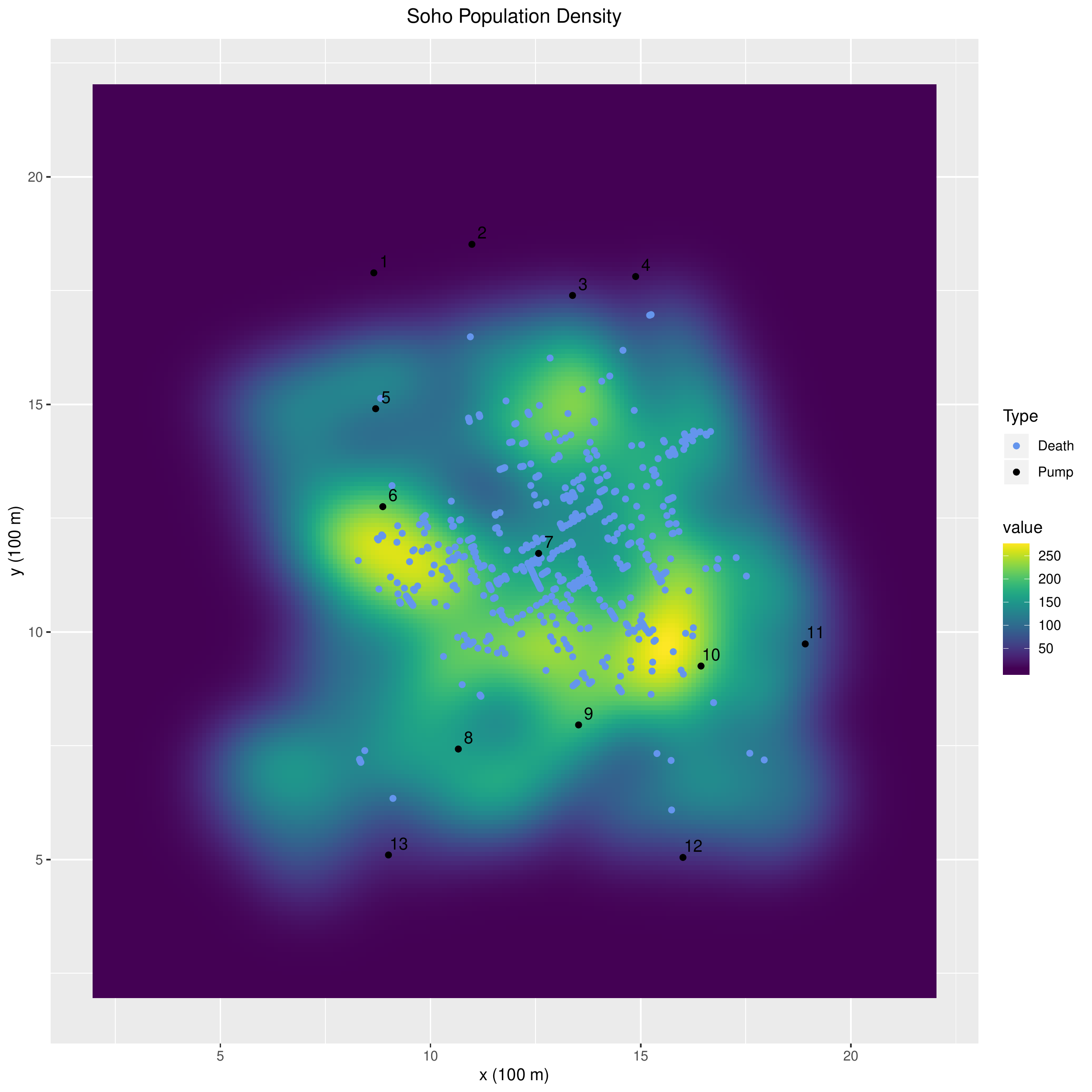}
		\caption{A kernel density estimate of the Soho, London population in 1854. Cholera death locations are plotted in blue as well as the water pumps numbered 1-13, with pump 7 being the Broad St. water pump. The estimated total population is 21,345. }
		\label{figure::Data Analysis Snow Population Density Plot}
	\end{figure}

	We now apply our proposed methodology to Dr. John Snow's cholera outbreak. The data set consists of $N=578$ observed cholera death locations. We fit the locations according to a LGCP with intensity $\lambda(\bs{s})$ as in \eqref{Eqn::LGCP Log Lambda Intensity} with the region of Soho, London represented by $\Omega = [200 \text{ m},2,200 \text{ m}]^2$. A priori, we assume $\bsbeta \sim N(\bs{0},2\text{I})$, where $\bsbeta$ consists of an intercept $\beta_0$ and the distance to the Broad St. water pump (the source of the outbreak) $\beta_1$. The estimated population kernel density estimate (see Figure \ref{figure::Data Analysis Snow Population Density Plot}) $\log{ \left( pd(\bs{s}) \right) }$ serves as the offset. The prior choices for the variance and spatial scale $\left( \kappa, \xi \right)$ of the spatial random field are detailed in Appendix \ref{Appendix::Priors}.    
	
	We fit the LGCP model via MCMC following the technique described in Appendix \ref{Appendix::Approximate Inference for LGCPs}. We tune our Metropolis-Hastings sampler according to the adaptive tuning scheme of \cite{roberts2009examples} for 250,000 burn-in samples. Another 250,000 post burn-in samples are stored for inference. The resulting parameter estimates, including the fixed effects, effective range, and marginal variance of the spatial random field are summarized in Table \ref{table::Data Analysis Params}. We note that $\hat{\beta}_1 = -0.946$ suggests that the further an individual lives from the Broad St. pump, the less susceptible they are to cholera death.   
	
\begin{figure}[H]
		\centering
		\includegraphics{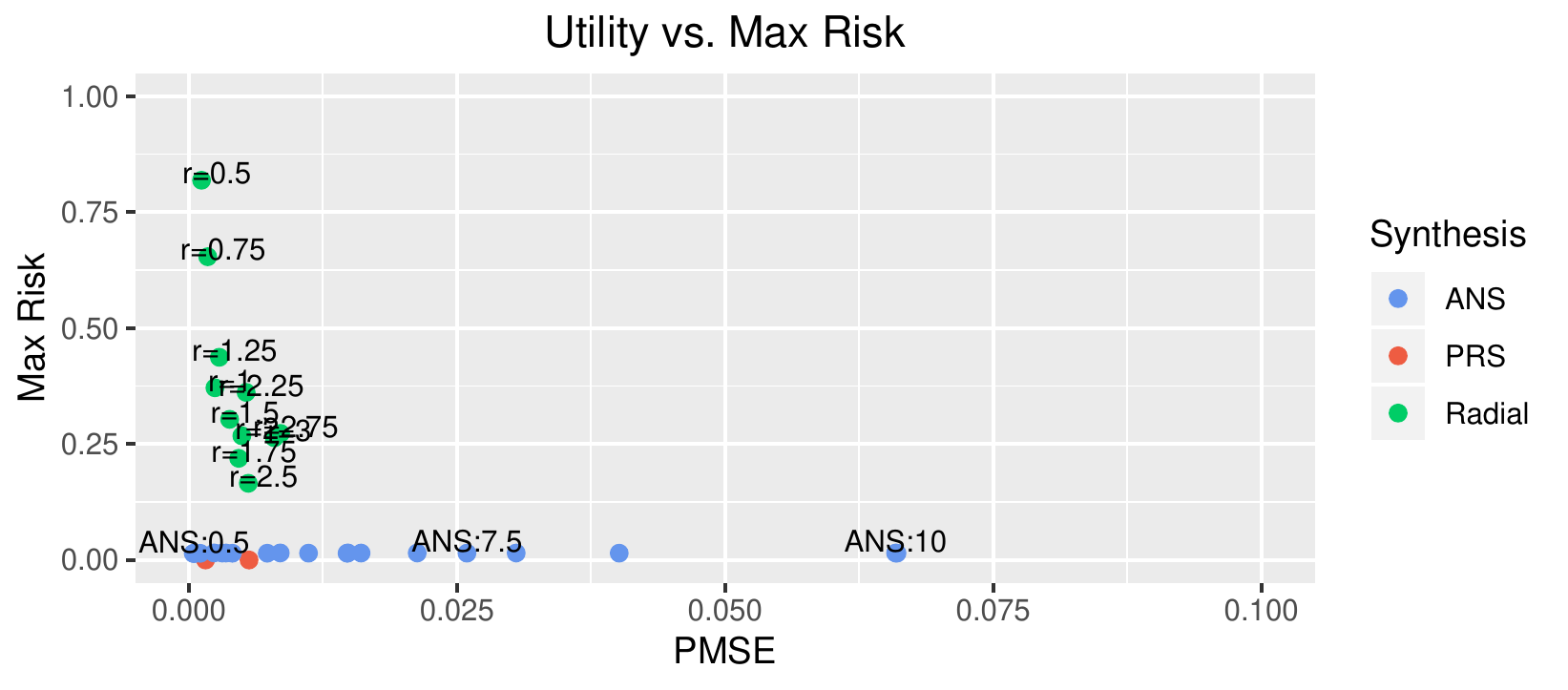}
		\caption{Plot of max disclosure risk vs. \textit{pMSE} for \textit{ANS}, radial synthesis, and \textit{PRS} datasets. Note that only the min and max utility are plotted for \textit{PRS}.}
		\label{figure::Data Analysis Risk vs. Utility}
	\end{figure}

	Using the fitted model, we generate 15 synthetic datasets via \textit{PRS} following Section \ref{section::Posterior Resampling Synthesis}. Additionally, 20 \textit{ANS} datasets are generated with noise levels $\sigma^2 = 0.5,1,1.5,...,9.5,10$ following Section \ref{section::Additive Noise Synthesis}. 10 synthetic sets via radial synthesis with radii $r  =$ 50 m, 100 m, ... , 300 m as detailed in Section \ref{section::Radial Synthesis}. Metropolis-Hastings samplers are used to assess utility and disclosure risks following Sections \ref{section::Evaluating Utility} and \ref{section::Evaluating Disclosure Risks}. Each sampler drew 500,000 samples, with 250,000 discarded as burn-in. The results are summarized in Figure \ref{figure::Data Analysis Risk vs. Utility}.
	
	\begin{figure}[H]
		\centering
		\includegraphics{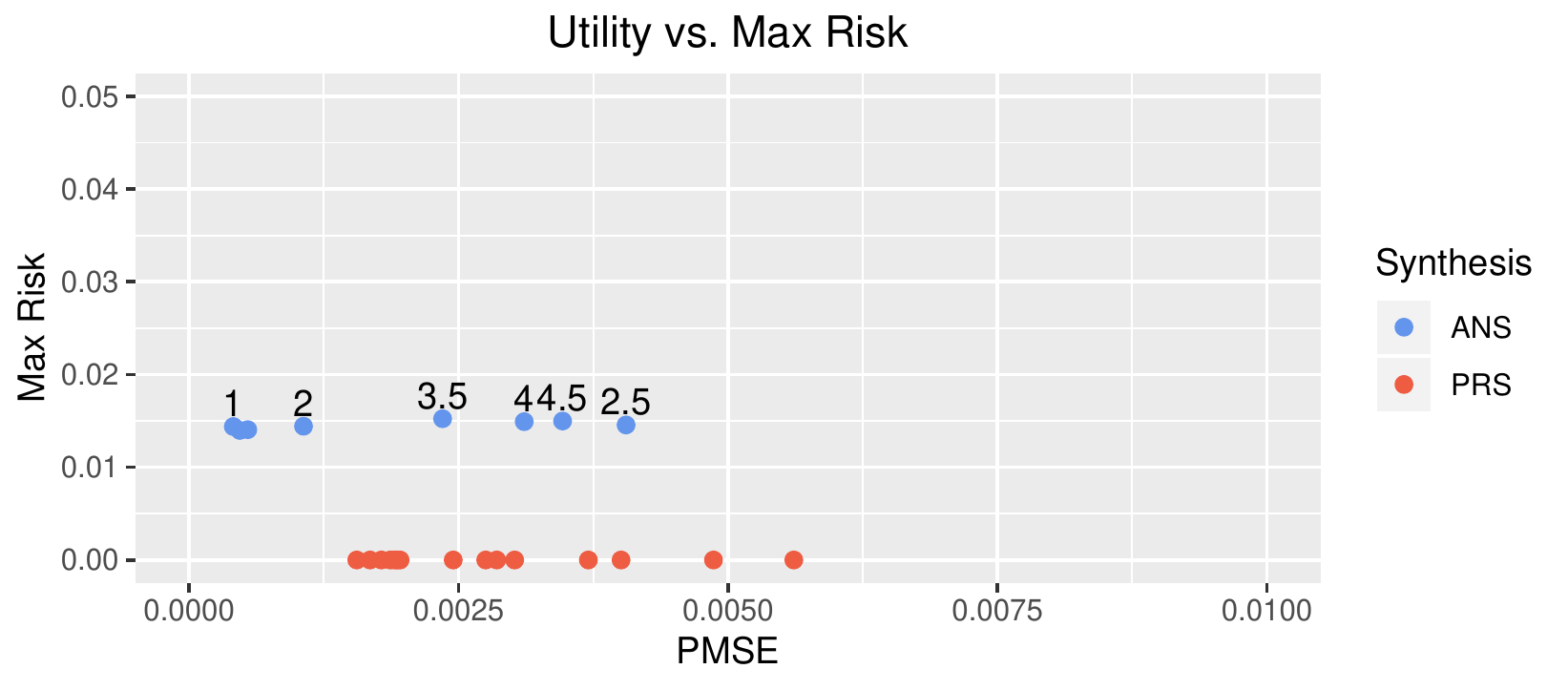}
		\caption{Plot of max disclosure risk vs. \textit{pMSE} for \textit{ANS} and \textit{PRS} datasets with max disclosure risks all less than 0.005. The noise level $\sigma^2$ is displayed above each \textit{ANS} point. All 15 \textit{PRS} data sets are plotted.}
		\label{figure::Data Analysis Risk vs. Utility_Zoom}
	\end{figure}
	
	From Figure \ref{figure::Data Analysis Risk vs. Utility} it is clear that synthetic sets generated according to \textit{PRS} offer the lowest max disclosure risks. In Figure \ref{figure::Data Analysis Risk vs. Utility}, we only plot the max and min utility scores for \textit{PRS}, as the max disclosure risk for all considered \textit{PRS} datasets was 6.28e-10. Figure \ref{figure::Data Analysis Risk vs. Utility} clearly shows that max disclosure risks for data sets generated from radial synthesis are consistently larger than \textit{ANS} and \textit{PRS} datasets. In Figure \ref{figure::Data Analysis Risk vs. Utility_Zoom}, we see that \textit{ANS} data sets with small noise levels offer an improvement in data utility with an increase in max disclosure risk in comparison to \textit{PRS}. For large noise values, the \textit{ANS} datasets offer small reductions in max disclosure risk at the expense of data utility. 
	
\begin{figure}[H]
		\centering
		\includegraphics[scale=0.5]{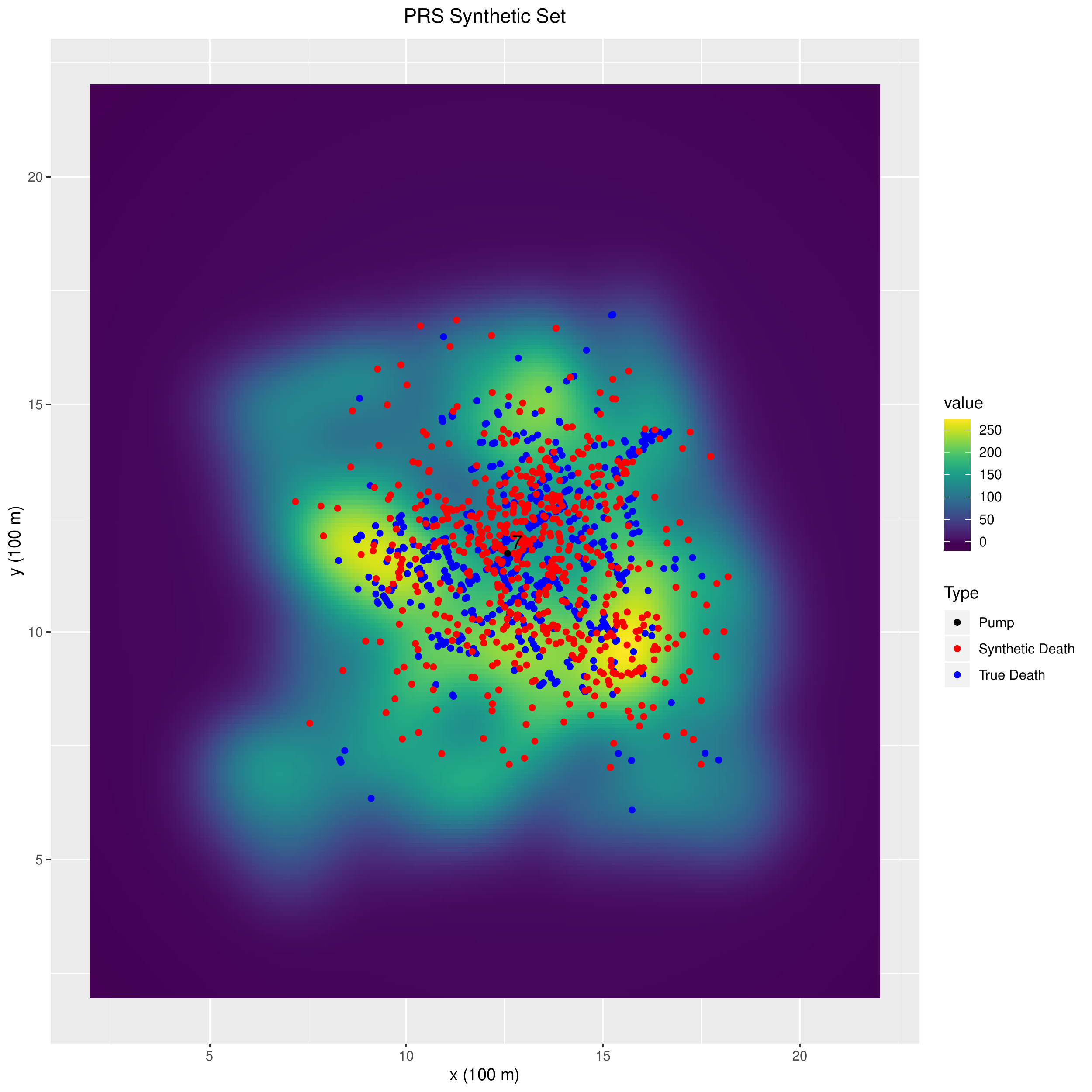}
		\caption{A plot of the fitted intensity surface for the optimal \textit{PRS} data set with max disclosure risk of 1.205e-17 and \textit{pMSE} score of 0.0016. The true deaths (blue), \textit{PRS} synthetic locations (red), and the Broad St. pump (black) are plotted as well.  }
		\label{figure::Data Analysis PRS_Best}
	\end{figure}

	The goal of this work is to generate a synthetic data set that offers low disclosure risks while preserving scientific inference. Since the max disclosure risk for the 15 considered \textit{PRS} data sets were all less than 6.28e-10, we suggest releasing the \textit{PRS} data set that offers the highest data utility with a \textit{pMSE} score of 0.0016. Recall that a small \textit{pMSE} score corresponds to higher data utility. The synthetic intensity surface, synthetic locations, and confidential locations are plotted in Figure \ref{figure::Data Analysis PRS_Best}. In Table \ref{table::Data Analysis Params}, we see that the \textit{PRS} fixed effect estimates, effective range, and marginal variance are similar to the confidential set (see Table \ref{table::Data Analysis Params}). In turn, Dr. Snow could still have determined the source of the cholera outbreak from the \textit{PRS} data set.  
	
	\begin{table}[H]
		\center 
		\begin{tabular}{lllll}
			Parameter &  Posterior Mean Est. &  95\% CI & \textit{PRS} Posterior Mean Est. & \textit{PRS} 95\% CI \\
			\hline
			$\beta_0$ & -0.771 & (-1.928, 0.387) & -0.799 & (-1.867, 0.431) \\ 
			$\beta_1$ & -0.946 & (-1.183,-0.729) & -0.829 & (-1.046, -0.642) \\ 
			$\rho$ & 3.487 & (2.027, 8.065) & 6.506 & (3.831, 9.177)   \\ 
			$\tau$ & 0.725 & (0.241, 2.577) & 0.554 & (0.269,1.143)  \\
		\end{tabular}
		\caption{Posterior mean estimates and corresponding 95\% credible intervals (CI) for the effective range and marginal variance of the spatial random field for the confidential and \textit{PRS} data sets.}
		\label{table::Data Analysis Params}
	\end{table}

	The goal of this work was to produce a synthetic data set that reduces individual disclosure risks while preserving data utility. In this case study, we observed that even for large radial perturbations (300 m), we could always find a noise level for \textit{ANS} which offers improved data utility and reduced disclosure risks. Similarly, \textit{PRS} datasets always offer reduced disclosure risks and improved data utility in comparison to radial synthesis. In summary, both of our proposed synthesis methods out perform radial perturbation in terms of risk vs. utility. Since the \textit{PRS} data sets considered offered lower max disclosure risks than any of the radial perturbation or \textit{ANS} data sets, we suggested the dissemination of the \textit{PRS} data set that offered the highest data utility. We then showed that the chosen \textit{PRS} data set offered inference similar to the confidential data set.  

	\section{Discussion} \label{section::Discussion}
	In this work we proposed two novel Bayesian approaches for generating fully-synthetic location-only datasets. We introduced a novel risk metric for a high intrusion scenario. We demonstrated that CPO estimates can easily be obtained for spatial location data generated from LGCPs, and allow for computationally efficient approximations of individual disclosure risks for each synthesis method. We adapted the \textit{pMSE} statistic for spatial point process data generated from LGCPs to evaluate data utility. We performed a case study of Dr. John Snow's cholera outbreak data set; showing that \textit{PRS} and \textit{ANS} offer an improved risk to utility ratio in comparison to radial perturbation. 
	
	In this work, we have conducted a Bayesian analysis of risk vs. utility for radial synthesis. Among other common methods used to privatize disease case locations prior to data dissemination are data aggregation and suppression \citep{armstrong1999geographically,wang2012multiple}. Spatial aggregations attempt to reduce disclosure risks by coarsening the geographic scale; e.g., reporting disease incidences at the county scale. However, any coarsening of the geographic scale diminishes the level at which a spatial analysis can be performed, and localized geographic trends/hazards can easily be missed \citep{armstrong1999geographically}. Data suppression attempts to reduce disclosure risks by removing high risk individuals, such as spatial outliers, from a dataset. If many individuals are deemed high risk, many locations may need to be suppressed, limiting the quality of any spatial analysis. In this work, we chose to focus on generating spatial disease case location data sets that contained the exact number of observations as the confidential set. Radial perturbation produces data sets consisting of point referenced disease cases with the same amount of observations as the original data set. For this reason, radial perturbation was considered as the baseline case, while aggregation and suppression were not.
	
	In this work, we considered an intrusion scenario in which an intruder attempts to identify a confidential location within a radius \textit{r} of the truth given 1) the synthetic dataset, 2) knowledge of the synthesis method, and 3) identification of all but the last confidential location. \cite{quick2015bayesian} considered a disclosure risk metric that defines an individual to be at high risk if they are spatially close to other individuals with similar attributes. Clearly this metric is not suited to handle location-only data. \cite{wang2012multiple} consider an intrusion case similar to ours. Their risk metric first computes the expected euclidean distance between an intruder's guess of the confidential geography with respect to $\pi{\left( \bs{s}_k | \mathcal{S}_{-k},\mathcal{S}^{\dagger} \right) }$, denoted R1. The number of actual cases within a radius of R1 of the truth are then counted as a measure of reasonable guesses for the confidential location. Our disclosure risk metric computes the probability that an individual is uniquely identified within a pre-specified radius \textit{r} of the truth. Our metric also contains population density as an offset, allowing the metric to account for different risk levels in regions of dense and sparse populations. 
	
	We proposed a utility metric based on the \textit{pMSE} statistic. Previous work by \cite{quick2015bayesian} assessed data utility with the K-function. The K function is a measure of spatial dependence that computes the expected number of events within a radius \textit{h} for each observation in the data set \citep{bartlett1964spectral,ripley1976second}. An \textit{ANS} generated data set will have a dependence structure with similar spatial scale and marginal variance that differs from the confidential set by some user defined noise level. \textit{PRS} data sets have the same spatial scale and marginal variance as the confidential set. Both of our proposed synthesis methods produce data sets with correlative structures similar to the confidential data set. In turn, they will likely produce similar K-function estimates. The \textit{pMSE} is a data-based utility metric that determines the utility of a synthetic data set based on how well a synthetic set emulates the true data set. We tailored the predicted probabilities used to classify conditional data in the \textit{pMSE} to LGCPs. In turn our metric accounts for the spatial dependence structure of the synthetic data set as well. We elected to use the \textit{pMSE} due to its intuitive interpretation; as the \textit{pMSE} can be viewed as the mean square misclassification score between synthetic and confidential locations.   
	
	%Wang and Reiter \cite{wang2012multiple} suggested sampling synthetic locations from approximate conditional distributions of each location given non-synthesized attributes. 	Quick et al. \cite{quick2015bayesian} suggested generating fully-synthetic locations with attributes by modeling each individual's location conditioned on their attributes according to a LGCP. The approach of Quick et al. \cite{quick2015bayesian} preserves the local spatial dependence structures of the confidential data. In a similar fashion, we have elected to model the confidential dataset according to a LGCP. We note that both the works of Wang and Reiter \cite{wang2012multiple} and Quick et al. \cite{quick2015bayesian} suggested generating synthetic locations by sampling from the fitted model. In our work, we have proposed two methods that first randomly alter the summary statistic ($\lambda$) prior to generating synthetic locations. 
	
	In summary we have developed two novel methods for generating fully-synthetic location-only data. We demonstrated that CPOs are easily obtainable for spatial point process data generated by LGCPs. We proposed a disclosure risk metric and model based data utility metric suited for synthetic location data with no attributes that could easily be evaluated with CPO estimates. We showed that our proposed methodology outperforms the common approach of radial synthesis in a case study of Dr. John Snow's cholera outbreak dataset. We illustrated that \textit{PRS} offers small max disclosure risk while preserving data utility. Our second proposed synthesis method \textit{ANS} was shown to offer the best utility for small noise levels. For this reason we believe both of our proposed synthesis methods should be used with the goal of balancing the risk vs data utility trade off.   

%%%%%%%%%%%%%%%%%%%%%%%%%%%%%%%%%%%%%%%%%%%%%%%%%%%%%%%%%%
%%%%%%%%%%% Bib %%%%%%%%%%%%%%%%%%%%%%%%%%%%%%%%%%%%%%%%%%
%%%%%%%%%%%%%%%%%%%%%%%%%%%%%%%%%%%%%%%%%%%%%%%%%%%%%%%%%%
\bibliographystyle{Chicago}
\bibliography{Snow_Paper_bib}

%%%%%%%%%%%%%%%%%%%%%%%%%%%%%%%%%%%%%%%%%%%%%%%%%%%%%%%%%%
%%%%% Appendix %%%%%%%%%%%%%%%%%%%%%%%%%%%%%%%%%%%%%%%%%%%
%%%%%%%%%%%%%%%%%%%%%%%%%%%%%%%%%%%%%%%%%%%%%%%%%%%%%%%%%%
	\begin{appendices}
	\section{Appendix}
	
	\subsection{Finite Element Approximations for Mat\'ern GRFs} \label{Appendix::Finite Element Approximations for Matern GRFs}
	
	Stationary Mat\'ern random fields arise as stationary solutions to the stochastic partial differential equation 
	\bea 
		\left(\kappasq - \triangle\right)\eta(\bs{s}) = \xi \mathcal{W}(\bs{s}), \quad \bs{s}\in \mathbb{R}^2, \label{Eqn::SPDE}
	\eea  
	where $\triangle = \sum_{i=1}^{2}\frac{\partial^2}{\partial s_i^2}$ is the Laplacian operator in 2-dimensions, $\kappa>0$ is a spatial scale parameter, $\tau>0$ is a variance parameter, and $\mathcal{W}(\bs{s})$ is Gaussian white noise. Whittle showed that solutions to \ref{Eqn::SPDE} are GRF with Mat\'ern covariance given by 
	\bea 
		C(\bs{u},\bs{v}) = \xi^2 (\kappa||\bs{v}-\bs{u}||)K_1(\kappa||\bs{v}-\bs{u}||), \quad \bs{v},\bs{u} \in \mathbb{R}^2, \label{Eqn::Matern Covariance}
	\eea  
	where $||\cdot||$ denotes Euclidean distance, and $K_1(\cdot)$ is an order one Bessel function of the second kind \citep{whittle1954stationary,whittle1963prediction}. The marginal variance is given by $\sigma^2 = \xi^2 / \left(4\pi \kappa^2\right)$. The approximation of the effective range is given by $\rho = \sqrt{8}/\kappa$ \citep{lindgren2011explicit}. 
	
	Though the analytic solution provides useful insights, model fitting and parameter estimation are often facilitated by considering a numerical approximation. \cite{lindgren2011explicit} proposed the use of a finite element approximation to the stochastic weak formulation of the SPDE $(\kappasq - \triangle)^{\alpha / 2}\eta(\psi) = \xi \mathcal{W}(\psi)$, where $\{\psi\}$ is a set of test functions. The finite element method (FEM) solution begins by expressing the solution, $\eta(\bs{u})$, as a basis expansion 
	\bea 
		\eta(\bs{s}) = \sum_{i=1}^{n} \phi_i(\bs{s}) w_i, \quad \bs{s} \in \Omega, \label{Eqn::Basis expansion}
	\eea  
	where $\{\phi_i(\bs{s})\}_{i=1}^{n}$ is a set of basis functions on $\Omega$. The solution is only required to hold for a finite collection of $\psi_i$. The Galerkin method approximate solution is obtained by setting $\{\psi_i\}_{i=1}^{n} = \{\phi_i\}_{i=1}^{n}$. 
	
	\cite{lindgren2011explicit} formulated an FEM approximation by considering $\{\phi_i(\bs{s})\}_{i=1}^{n}$ to be piecewise triangular basis functions.  The basis functions are constructed by partitioning the spatial region of interest, $\Omega \subset \mathbb{R}^d$, into non-overlapping triangular regions. The corners of the triangles, referred to as vertices, are assigned $n$ Gaussian weights, denoted $w_i$. Each $\phi_i$ is defined to be 1 at vertex \textit{i} and 0 at all other vertices. \cite{lindgren2011explicit} derived the distribution of the weights 
	\bea 
		\bs{w} | \xi,\kappa \sim \mathcal{N}\left( \bs{0}, \xi^2 \bs{Q}^{-1}_{\kappasq} \right), \label{Eqn::weight dist} 
	\eea 
	where $\bs{Q}^{-1}_{\kappasq} = \bs{L}^{-1}\bs{C}\bs{L}^{-1}$, with $\bs{L}=\kappasq \bs{C} + \bs{G}$. The matrices used to define \textbf{L} are given by $C_{ij} = \int_{\Omega} \phi_i(\bs{s}) d\bs{s}$, and $G_{ij} = \int_{\Omega} \nabla \phi_i(\bs{s}) \nabla \phi_j(\bs{s}) d\bs{s}$. Under lattice refinement, the FEM solution converges to the true solution (see Appendix C.5 of \cite{lindgren2011explicit}).

	\subsection{Approximate Inference for LGCPs} \label{Appendix::Approximate Inference for LGCPs} 
	The stochastic integral, $\int_{\Omega} \lambda(\bs{s})d\bs{s}$, in the likelihood of \eqref{Eqn::LGCP Cox Process Likelihood} cannot be computed analytically. Numerical integration schemes are used to approximate Cox process likelihoods. A common approach is to grid up the spatial domain into rectangular regions and evaluate the integral as a weighted sum \citep{diggle2013spatial}. Approximations are improved by refining the lattice into rectangles of smaller area. This becomes computationally burdensome, as rectangular lattice refinements often produce a mesh with fine partitions in regions containing few observations. 
	
	A desirable mesh possess a finer partition in regions with many observations, and a sparse partition in regions of few observations; a feature difficult to obtain for regularly spaced lattices. \cite{simpson2016going} proposed the use of a second order approximate dual-cell mesh, known as the Voronoi mesh, by joining the triangular elements required for fitting the Gaussian process described in Appendix \ref{Appendix::Finite Element Approximations for Matern GRFs} at their centroids. The integral can now be approximated by 
	\bea 
		\int_{\Omega} \lambda(\bs{s})d\bs{s} \approx  \sum_{i=1}^{n} \tilde{\alpha}_i \exp(\log(o(\tilde{\bs{s}}_i))+\bs{x}\p(\tilde{\bs{s}}_i)\bsbeta + \sum_{j=1}^{n}\phi_j(\tilde{\bs{s}}_i) ), \label{Eqn::Appendix Voronoi Dual Mesh Sum}
	\eea 
	where $\{\tilde{\bs{s}}_i\}_i^{n}$ are mesh nodes corresponding to the FEM mesh, and $\tilde{\alpha}_i$ is the volume of the $i^{th}$ dual cell produced by the Voronoi mesh. 
	
	To construct the approximate likelihood first define the projection matrix $P_{ij} = \phi_j(\bs{s}_i)$, and let the spatially continuous covariates at the observed locations and mesh nodes be denoted by $\bs{X} = [\bs{1}_{Nx1},\bs{x}(\bs{s}_1)...\bs{x}(\bs{s_N})]$, and $\tilde{\bs{X}} = [\bs{1}_{nx1}, \tilde{\bs{x}}(\bs{s}_1)...\tilde{\bs{x}}(\bs{s_n})]$ respectively. We then define  $\log(\bs{\eta}) = (\bs{w}\p + \bsbeta\bsX\p, \bs{w}\p\bs{P}\p + \bsbeta\p\tilde{\bs{X}})$, $\bs{\alpha} = (\tilde{\alpha},\bs{0}\p_{Nx1})$, and $\bs{y} = (\bs{0}\p_{Nx1},\bs{1}\p_{nx1})$. The approximate log-likelihood is given by 
	\bea 
		\log(\mathcal{S},N|\lambda) = \sum_{i=1}^{N+n} y_i\log(\eta_i) - \alpha_i\eta_i. \label{Eqn::Appendix Approximate LGCP Likelihood}
	\eea     
	We note that \eqref{Eqn::Appendix Approximate LGCP Likelihood} now resembles the sum of $N+n$ independent Poisson random variables with rate $y_i\alpha_i$. Model fitting can now be performed similarly to Poisson spatial generalized linear mixed models.
	
	\subsection{Circular Synthesis Disclosure Risk Details} \label{Appendix::Circular Synthesis Disclosure Risk Details}
	Here we derive the CPO estimate used to evaluate disclosure risks for radial synthesis given in \eqref{Eqn::Radial Synthesis CPO} of Section \ref{section::Radial Synthesis Disclosure Risks}. First note that the joint likelihood can be written as follows,
	\bea 
		\pi{\left(\mathcal{S},\mathcal{S}^{\dagger}\right)} = \pi{\left(\mathcal{S}^{\dagger}|\mathcal{S} \right)}\pi{\left(\mathcal{S}\right)} = \left(\frac{1}{\pi r^2}\right)^{N} \left( \prod_{k=1}^{N} \mathbb{I}_{\left\{||\bs{s}_k-\bs{s}||<r\right\} }(\bs{s}_k^{\dagger}) \right)  \pi{\left(\mathcal{S}\right)}. \label{Eqn::Appendix Circular S_dagger given S} 
	\eea 
	Observe that the conditional density for $\mathcal{S}_{-k}$ and $\mathcal{S}$ given $\lambda$ is proportional to $\pi{ \left( \mathcal{S}^{\dagger},\mathcal{S}|\lambda \right) }$. 
	\bea 
		\pi{ \left( \mathcal{S}_{-k}, \mathcal{S}^{\dagger} | \lambda \right) } &=& \int_{\Omega} \pi{ \left( \mathcal{S}^{\dagger} | \mathcal{S} \right) }\pi{ \left( \mathcal{S} | \lambda \right) } d\bs{s}_k \nt \\
		&=& \left(\frac{1}{\pi r^2}\right)^{N} \frac{\prod_{i \neq k} \lambda(\bs{s}_i)}{N!\exp{\left( \int_{\Omega} \lambda(\bs{s}) d\bs{s} \right)}} \int_{\Omega} \mathbb{I}_{\left\{||\bs{s}_k-\bs{s}||<r\right\} }(\bs{s}_k^{\dagger}) \lambda(\bs{s}_k)d\bs{s}_k \nt \\
		&=& \left(\frac{1}{\pi r^2}\right)^{N} \frac{\prod_{i \neq k} \lambda(\bs{s}_i)}{N!\exp{\left( \int_{\Omega} \lambda(\bs{s}) d\bs{s} \right)}} \int_{\mathcal{B}_r(\bs{s}_k^{\dagger})} \lambda(\bs{s})d\bs{s} \nt \\
		&=& \left( \frac{\lambda(\bs{s}_k)}{\int_{\mathcal{B}_r\left(\bs{s}_k^{\dagger}\right)} \lambda(\bs{s}) d\bs{s} } \right) \pi{ \left( \mathcal{S}, \mathcal{S}^{\dagger} | \lambda \right) }. \label{Eqn::Appendix Circular S_k S_dagger given Theta Joint} 
	\eea 
	We use \eqref{Eqn::Appendix Circular S_dagger given S} and \eqref{Eqn::Appendix Circular S_k S_dagger given Theta Joint}  to obtain the CPO estimate
	\bea 
		\pi{ \left( \bs{s}_k | \mathcal{S}_{-k}, \mathcal{S}^{\dagger} \right) } = \left[  \frac{ \pi{ \left( \mathcal{S}_{-k},\mathcal{S}^{\dagger} \right) } }{ \left( \mathcal{S},\mathcal{S}^{\dagger} \right) } \right]^{-1} &=& \left[  \frac{ \int \pi{ \left( \mathcal{S}_{-k},\mathcal{S}^{\dagger} | \lambda \right) } \pi{ (\lambda) d\lambda } }{ \pi{ \left( \mathcal{S},\mathcal{S}^{\dagger} \right)} } \right]^{-1} \nt \\ 
		&=& \left[  \frac{ \int \left( \frac{\int_{\mathcal{B}_r\left(\bs{s}_k^{\dagger}\right)} \lambda(\bs{s}) d\bs{s} }{\lambda(\bs{s}_k)} \right) \pi{ \left( \mathcal{S}, \mathcal{S}^{\dagger} | \lambda \right) } \pi{ (\lambda) d\lambda } }{ \pi{ \left( \mathcal{S},\mathcal{S}^{\dagger} \right)} } \right]^{-1} \nt \\ 
		&=& \left[  \frac{ \int \left( \frac{\int_{\mathcal{B}_r\left(\bs{s}_k^{\dagger}\right)} \lambda(\bs{s}) d\bs{s} }{\lambda(\bs{s}_k)} \right) \pi{\left(\mathcal{S}^{\dagger}|\mathcal{S} \right)} \pi{ \left( \mathcal{S} | \lambda \right) } \pi{ (\lambda) d\lambda } }{ \pi{\left(\mathcal{S}^{\dagger}|\mathcal{S} \right)} \pi{ \left( \mathcal{S} \right)} } \right]^{-1} \nt \\
		&=& \left[  \int \left( \frac{\int_{\mathcal{B}_r\left(\bs{s}_k^{\dagger}\right)} \lambda(\bs{s}) d\bs{s} }{\lambda(\bs{s}_k)} \right) \pi{ \left( \lambda | \mathcal{S} \right) } d\lambda  \right]^{-1} \nt  \\
		&=& \left[ \mathbb{E}_{\pi{ \left( \lambda | \mathcal{S} \right) }}\left[ \left( \frac{\int_{\mathcal{B}_r\left(\bs{s}_k^{\dagger}\right)} \lambda(\bs{s}) d\bs{s} }{\lambda(\bs{s}_k)} \right)  \right] \right]^{-1}. \label{Eqn::Appendix Circular CPO Estimate}  
	\eea 
	
	\subsection{Quadrature Scheme For Circular Domains} \label{Appendix::Quadrature Scheme For Circular Domains}
	Here we detail the quadrature scheme for numerical integration a over circular domain used to obtain disclosure risk estimates in Section \ref{section::Evaluating Disclosure Risks}. Let $\bs{s}_k = (x_k,y_k)$ represent a location such that $x_k,y_k > 0$. We wish to integrate 
	\bea 
		\int_{ \mathcal{B}_{r}(\bs{s}_k) } \lambda(x,y) dxdy = \int_ {-r}^{r} \int_{x_k - g(y)}^{x_k+g(y)} \lambda(x,y) dxdy \label{Eqn::Appendix Ball Integral}
	\eea 
	where $g(y) = \sqrt{(r^2 - (y-y_k)^2)}$. Making the change of variables $\tilde{y} = \frac{(y-y_k)}{r}$ and $\tilde{x} = \frac{(x-x_k)}{g(r\tilde{y}+y_k)}$, the integral in \eqref{Eqn::Appendix Ball Integral} becomes
	\bea 
		\int_{-1}^{1} \int_{-1}^{1} \lambda(g(r\tilde{y}+y_k)\tilde{x} + x_k, r\tilde{y}+y_k)*r*g(r\tilde{y} + y_k) d\tilde{x}d\tilde{y}. \label{Eqn::Appendix Integral Over Unit Square} 
	\eea   
	Notice that the integral in \eqref{Eqn::Appendix Integral Over Unit Square} is now computed over the square $[-1,1]^2$. We now partition $[-1,1]^2$ into $M$ equally sized squares of area $A_{xy}$. Let $(\tilde{x}_m,\tilde{y}_m)$ represent the center of each of the $M$ squares. The integral in \eqref{Eqn::Appendix Integral Over Unit Square} is now approximated by the sum 
	\bea 
		\sum_{m=1}^{M} \lambda(g(r\tilde{y}_m + y_k )\tilde{x}_m + x_k, r\tilde{y}_m + y_k )*r*g(r\tilde{y}_m + y_k)*A_{xy}. \label{Eqn::Appendix Quadrature Sum} 
	\eea  
	
	\subsection{Prior Choice for $(\kappa,\xi)$} \label{Appendix::Priors}
	Here we summarize the prior choice for $(\kappasq, \xi)$ suggested by \cite{lindgren2015bayesian}. Following the Appendix \ref{Appendix::Finite Element Approximations for Matern GRFs} we obtain an approximation to a Gaussian random field with Mat\'ern covariance. The basis expansion weights have variance and spatial scale hyperparameters $\xi^2$ and $\kappasq$ (see Appendix \ref{Appendix::Finite Element Approximations for Matern GRFs} equation \eqref{Eqn::weight dist}). The covariance is summarized by these parameters relationship to the spatial range $\rho$ and marginal variance $\sigma^2$ given by
	\bea 
		\rho = \frac{\sqrt{8}}{\kappa}, \quad \sigma^2 = \frac{ \xi^2 }{ 4\pi \kappasq}. \label{Eqn::Appendix Priors rho sigmasq} 
	\eea 
	From \eqref{Eqn::Appendix Priors rho sigmasq} it is clear that the marginal variance is influenced by both $\kappa$ and $\xi$. We would like a prior that captures this relationship. To do so, assume $\left( \theta_1,\theta_2 \right)  \sim N(\bs{0}, \Sigma_{\theta})$. Suppose we want the parameterization 
	\bea 
		\log(\rho) &=& \log(\rho_0) + \theta_1 \label{Eqn::Appendix Priors log_rho} \\ 
		\log(\sigma) &=& \log(\sigma_0) + \theta_2, \label{Eqn::Appendix Priors log_sigma_0} 
	\eea 
	where $\log(\rho_0)$ and $\log(\sigma_0)$ are the baseline range an marginal standard deviation. Using equations \eqref{Eqn::Appendix Priors rho sigmasq} and \eqref{Eqn::Appendix Priors log_rho} we can write 
	\bea 
		\log(\kappa) = \frac{1}{2}\log(8) - \log(\rho_0) - \theta_1 = \log(\kappa_0) - \theta_1, \label{Eqn::Appendix Priors log_kappa}
	\eea 
	where $\log(\kappa_0) = \frac{1}{2}\log(8) - \log(\rho_0)$. It follows from equations \eqref{Eqn::Appendix Priors rho sigmasq} and \eqref{Eqn::Appendix Priors log_sigma_0} that
	\bea 
		\log(\phi) &=& \log(\sigma_0) - \frac{1}{2}\log(4\pi) - \log(\kappa_0) + \theta_2 - \theta_1 \label{Eqn::Appendix Priors log_phi} \\
		&=& \log(\phi_0) + \theta_2 - \theta_1, 
	\eea 
	where $\log(\phi_0) = \log(\sigma_0) - \frac{1}{2}\log(4\pi) - \log(\kappa_0)$. Equations \eqref{Eqn::Appendix Priors log_kappa} and \eqref{Eqn::Appendix Priors log_phi} now give a joint prior on $(\kappa,\xi)$ that captures their dependent influence on the marginal variance. In Section \eqref{section::Data Analysis} we took $\log{\left( \rho_0 \right) } = 0$ and $\log{\left( \sigma_0 \right)} = 0$.

\end{appendices}

\end{document}